\documentclass[oneside,preprint]{aastex}
\usepackage[T1]{fontenc}
\usepackage[latin9]{inputenc}
\usepackage{hyperref}
\hypersetup{urlcolor=blue, colorlinks=false}  
\usepackage{amstext}
\usepackage{amssymb}
\usepackage{graphicx}
\usepackage{lscape}
\usepackage{longtable}
\usepackage{verbatim}  
\usepackage{pdflscape}
\usepackage{multicol}
\usepackage{multirow}
\usepackage{graphicx}
\usepackage{natbib}
\makeatletter

\usepackage[english]{babel}
\usepackage{amsmath}

\makeatother

\shorttitle{Disk-outflow models in high mass star formation.}
\shortauthors{H. Farmer}
\begin{document}

\title{ Disk-outflow models as applied to high mass star forming
regions through methanol and water maser observations.}

\author{Hontas F. Farmer\altaffilmark{1}}
\affil{Department of Physics
Northern Illinois University}
\affil{214 Faraday Hall DeKalb, IL 60115}
\email{hfarmer@niu.edu}

\altaffiltext{1}{Department of Physics DePaul University, Chicago, Illinois, USA.}

\begin{abstract}
As the recent publication by Breen et al (2013) found Class II methanol masers are exclusively associated with high mass star forming regions. Based on the positions of the Class I and II methanol and
H$_{2}$O masers, UC H~II regions and 4.5~$\mu$m infrared sources,
and the center velocities ($v_{\text{LSR}}$) of the Class I methanol
and H$_{2}$O masers, compared to the $v_{\text{LSR}}$ of the Class
II methanol masers, we propose three disk-outflow models that may
be traced by methanol masers. In all three models, we have located
the Class II methanol maser near the protostar, and the Class I methanol
maser in the outflow, as is known from observations during the last
twenty years. In our first model, the H$_{2}$O masers trace the linear
extent of the outflow. In our second model, the H$_{2}$O masers are
located in a circumstellar disk. In our third model, the H$_{2}$O
masers are located in one or more outflows near the terminating shock
where the outflow impacts the ambient interstellar medium. Together,
these models reiterate the utility of coordinated high angular resolution
observations of high mass star forming regions in maser lines and
associated star formation tracers. 
\end{abstract}

\clearpage{}

\section{Introduction}

\label{Chapter1}

Understanding the process of the formation of high mass stars with
masses about eight times that of our Sun, or greater, remains a challenge
in modern astronomy. High mass stars are important because they inject
matter and energy into the interstellar medium of galaxies throughout
their lifetimes. Observing high mass star formation is difficult,
though, because massive star forming regions are farther away than
their low mass counterparts, high mass stars form in clusters, and
begin the process of nuclear fusion even as they are accreting matter.
One of the key issues is whether high mass star formation is a scaled-up
version of the low mass star formation process, complete with disk
and outflows, or whether high mass stars form by coalescence of low
mass stars. Being compact and bright sources, masers offer a unique
opportunity to peer into the deep interior of high mass star forming
regions at high angular resolution. In this paper, we use data on
methanol and water masers from the literature, along with other associated
star formation tracers, to gain insight into the high mass star formation
process by investigating whether disk-outflow systems are compatible
with these maser data.  The exclusive association of Class II methanol masers with high mass protostars was shown by \citet{Breen2013}.  In the thesis \citet{Farmer2013} a similar analysis was carried out for a larger selection of regions.  This paper is intended as a journal publishable write up of \citet{Farmer2013} updated with Breen's results.  In this section, we begin with a discussion of low mass star formation in \S\ \ref{LMSF}, high mass star formation in \S\ \ref{HMSF}, and masers and maser observations
in \S\ \ref{maserobservations}.

\subsection{Low mass star formation\label{LMSF}}

Over the last fifty years, astronomers have gained a fairly good understanding
of the basic process of low mass star formation. Star formation is
initiated when dense cores inside molecular clouds become gravitationally
unstable and collapse. Depending on the size of the core, one or more
stars will be formed. Small cores may form a single star, larger cores
will fragment and form multiple stars (some authors use the term ``clumps''
for larger cores).

The exact details of the process of collapse are model dependent,
but the following are generally common to most models \citep{McKee2007},
Gravitational collapse builds up the central density in the core of
the cloud or cloud fragment. The energy generated by the gravitational
collapse is radiated away because the cloud remains optically thin
(i.e., transparent to radiation); as a result, during the initial
stages of the collapse, most models treat the cloud as isothermal.
The isothermal collapse phase produces a central condensation of matter
and ends with the formation of a stable object at the center called
a protostar, surrounded by a gaseous envelope. With the increasing
density in the gaseous envelope, it becomes harder for the energy
released by gravitational contraction to escape from the star, as
a result of which the internal temperature of the protostar rises.
The protostar then enters its main accretion phase, in which it builds
up its mass from a surrounding infalling envelope and accretion disk,
even as it continues to get hotter. Some of the accreted material
is ejected along opposite directions perpendicular to the accretion
disk, and this is known as a bipolar outflow (e.g., see Figure \ref{badartist}).
These outflows are believed to help in carrying away the excess angular
momentum of the infalling matter; this is necessary because of the
so-called ``angular momentum problem'' in star formation. This problem
arises because while rotation is not a significant source of support
in molecular clouds against gravity, a decrease of just a factor of
100 in size (while conserving angular momentum) increases the rotational
energy to the point where it can support the cloud against gravitational
collapse. Yet, since decreases by factors of $10^{5}$ or greater
in size are required for a core to become a star, if it were the case
that a decrease of just $10^{2}$ were enough to bring the rotational
energy into equilibrium with gravitational energy, then stars could
never form. Mechanisms have been sought for how to get rid of the
angular momentum during collapse, and bipolar outflows and the magnetic
fields that are present in these outflows are believed to carry away
the excess angular momentum, but the exact mechanism is still not
fully understood.

When the protostar has accumulated most of its final mass (that it
will have during its lifetime as a main sequence star) it is generally
known as a pre-main-sequence (PMS) star, although the use of the term
varies among authors (\citealt{McKee2007}). Such stars continue to
supply most of their luminosity (i.e., the energy output in, e.g.,
ergs s$^{-1}$) by contracting and releasing gravitational potential
energy, but now their luminosities and surface temperatures increase.
Eventually, the interior becomes hot enough to initiate hydrogen fusion,
and this takes over as the cause of the luminosity of the star. The
star soon comes into hydrostatic equilibrium between the outward thermal
pressure and the inward gravitational pull and reaches its main sequence
phase, in which it will spend 90\% of its life span, steadily fusing
hydrogen to helium.

\subsection{High mass star formation\label{HMSF}}

The process of high mass star formation is not as well understood
as low mass star formation. For a number of reasons high mass stars
constitute a very difficult observational and theoretical problem.
High mass star forming regions are located farther from us than low
mass star forming regions; the nearest low mass star forming region
is Ophiuchus at a distance of 120 pc (\citealt{Loinard2008}), whereas
the nearest high mass star forming region in Orion is 500 pc away
(\citealt{Genzel1981}); high mass stars usually form in clusters,
and because of their high mass, they begin fusing hydrogen even as
they are accreting material. It is difficult to find an isolated high
mass star in the process of forming with which to study the process.
If we observe an outflow, it is often difficult to figure out which
high mass protostar the outflow is coming from, because the protostars
are so close together. Moreover, high mass star forming regions are
obscured by more gas and dust than low mass star forming regions (\citealt{Zinnecker2007}).

Theoretical studies of this process must address how accretion can
continue in the face of the tremendous outward radiation pressure
generated by the onset of hydrogen fusion.

\subsection{Maser observations of high mass star forming regions\label{maserobservations}}
Observations of maser lines are ideal for studying high mass star
forming regions at high angular resolution, as masers are compact
and bright sources. There are two varieties of methanol maser, with Class II masers
exclusively associated with high mass star forming regions as shown by \citet{Breen2013} in 
their analysis of two regions ( G328.385+0131, G10.10+0.73 ). 

The division of methanol masers into two classes was originally based on association:
Class II masers are associated with protostars, and Class I masers
are found in outflows (\citealt{Menten1991}). Today, we know there
are differences in the pumping mechanism as well; Class II methanol
masers are believed to be radiatively pumped, while Class I masers
are pumped collisionally. This would also explain why Class II masers
are found near high mass protostars where they are pumped by the intense
radiation from the protostar (\citealt{Sobolev1997}), and why Class
I masers are found in outflows where they are pumped by collision
at the interfaces of shock fronts between the outflowing material
and the ambient interstellar material (\citealt{cragg92}). The most
well known Class II methanol masers occur at frequencies of 6.7 GHz
and 12.2 GHz, whereas Class I methanol masers occur at 44 GHz, 95
GHz, and 36 GHz.

\begin{figure}[h!]
\centering \includegraphics[width=1.2\textwidth]{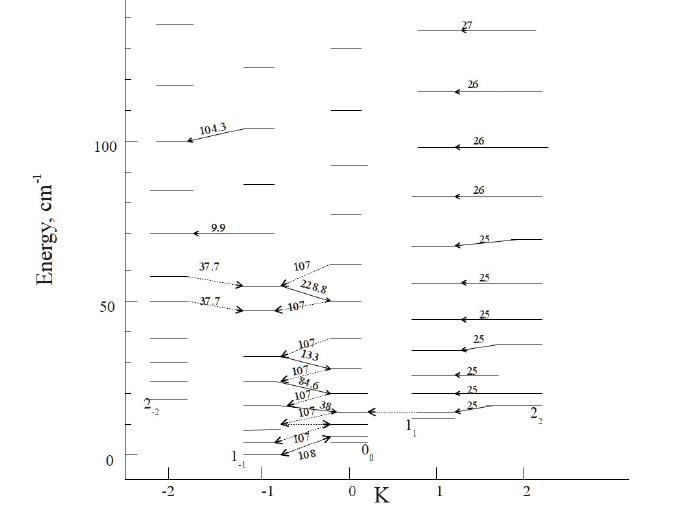}

\caption[Energy state diagram for E-methanol.]{Diagram showing the energy levels of type E methanol for different
values of K, where K is the projection of the total angular momentum
J onto the principal axes of the molecule. There are two species of
methanol, E and A, differentiated by the spin orientations of their
hydrogen atoms. This figure was derived from one due to \citet{Voronkov2013}
and used with the original authors permission.}

\label{methanol-state-diagram.} 
\end{figure}

\clearpage{}

\section{Data Sources and Collection}

\citet{Farmer2013} began by looking for high angular resolution data on Class I and
Class II methanol masers in star forming regions; angular resolution
is defined and discussed in the next paragraph. \citet{valtts2007}
is a compiled catalog of all the known Class I methanol masers. From
the 206 masers in their list, we picked the 44 GHz methanol masers
that had been observed at high angular resolution with the the Karl
G. Jansky Very Large Array (VLA); there were 32 sources of this kind.
For each of these 32 maser sources, we searched the literature for
Class II methanol masers within a 1$^{\prime}$ radius of the position
listed in the \citet{valtts2007} catalog. Since H$_{2}$O masers
are also signposts of star formation, we searched the literature for
H$_{2}$O masers near these Class I and II methanol masers. Recently,
\citet{cyganowski2011} have shown that 4.5 $\mu$m infrared sources
are associated with high mass star formation, so we also looked for
such sources near the Class I and II methanol masers. Finally, young
high mass stars are known to form ultra compact (UC) ionized hydrogen
(H II) regions by ionizing the hydrogen around them , so we also looked
for data on UC H II regions.

We wanted to work with the highest angular resolution data available
for our sources. Therefore, we preferred to use data taken with interferometric
telescopes like the VLA wherever available. The angular (or spatial) resolution
of a telescope is given by the ratio of the wavelength of observation
to the diameter of the telescope. When signals from two radio telescope
antennas are combined, the effective diameter of the telescope is
the distance between the antennas. Therefore, by locating telescopes
at large distances from each other as in the VLA, interferometers
achieve much greater resolution than a single dish radio telescope.
For example, the angular resolution of a single dish telescope of
diameter 25 m would be about 60$^{\prime\prime}$if it were observing at a frequency of 44 GHz. By way of comparison,
\citet{kurtz2004} were able to obtain a much higher angular resolution
of 5$^{\prime\prime}$, i.e., over ten times better, in their VLA
observations of 44 GHz methanol masers discussed below.

We will now discuss the characteristics of the observed data, specifically,
the papers we consulted, the telescope and observing parameters used
to observe these data, and any other relevant information.

\subsection{The Class I methanol maser data.}

The positions of the 44 GHz Class I methanol masers were taken from
the \citet{valtts2007} catalog. Most of the positions listed in this
catalog were taken from VLA observations reported in \citet{kurtz2004},
from which we obtained information on the center velocities ($v_{\text{LSR}}$)
and intensities of the methanol masers. \citet{kurtz2004} was a survey
of forty-four star forming regions in search of 44 GHz Class I methanol
masers with the goal of investigating the relationship between such
masers and shocked molecular gas. The angular resolution of this survey
was about 0.5$^{\prime\prime}$, and the velocity resolution was 0.17
km s$^{-1}$ (corresponding to a frequency resolution of about 24
kHz). The idea of frequency resolution comes from the finite bandwidth
of telescope receivers that must be sampled in discrete intervals.
So, for example, \citet{kurtz2004}'s survey used a frequency set
up in which the bandwidth was 3.125 MHz sampled with 128 channels,
to give a frequency resolution of 24 kHz. However, astronomers like
to work with velocities because they can be connected more easily
to the physical motions of the source(s) in the sky. For Galactic
observations such as those studied in this paper, one would convert
from one to the other using the non relativistic Doppler effect: $\Delta v/c=\Delta f/f$,
where $\Delta f$ is the shift in frequency from the observed frequency,
$f$. The root mean square (rms) noise in the maser spectra line profiles,
usually measured as one fifth of the peak-to-peak value in the line
observed in a source-free region of the sky, was typically 40 mJy
beam$^{-1}$, so they were able to detect masers down to a median
5-$\sigma$ limit of 0.2 Jy. For several
sources, \citet{kurtz2004} found evidence for a spatial correlation
between the 44 GHz masers and shocked molecular gas in agreement with
the view that these masers are produced by collisional pumping in
molecular outflows.

\subsection{The Class II methanol maser data}

The Class II methanol maser data were taken from a catalog of 6.7
GHz Class II methanol masers compiled by \citet{pestalozzi2005}.
This catalog contains 519 sources compiled from 62 references in the
literature. \citet{pestalozzi2005} found that Class II methanol masers
trace the molecular ring of our galaxy where massive OB star associations
are found, in agreement with the idea that methanol masers are clearly
associated with high mass star formation.

\citet{pestalozzi2005} relied heavily on \citet{caswell95}, which
was a survey of thirty-six sites of 6.7 GHz methanol masers and forty
sites of 1.7 GHz OH masers, conducted with the Australia Telescope
Compact Array (ATCA). The purpose of this survey was to study the
relationship between OH and CH$_{3}$OH masers and the evolution of
massive stars. The angular resolution of the methanol maser observations
was about 1.5$^{\prime\prime}$, the frequency resolution was 0.97
kHz, and the rms noise was 60 mJy beam$^{-1}$.

For more recent data in the literature, we used \citet{pandian2011}
and \citet{cyganowski2009}. \citet{pandian2011} observed 57 Class
II methanol masers at 6.7 GHz with the Multi-Element Radio-Linked
Interferometer Network (MERLIN, Jodrell Bank) with an angular resolution
of 60 milliarcseconds (mas), a frequency resolution of 3 kHz ($\equiv0.13$
km s$^{-1}$), and an rms noise of about 35 mJy beam$^{-1}$. They
found very close correspondence between methanol masers and 24 $\mu$m
mid-infrared sources, lending further support to theoretical models
that predict methanol masers are pumped by infrared dust emission
in the vicinity of high-mass protostars. \citet{cyganowski2009} observed
44 GHz Class I and 6.7 GHz Class II methanol masers in 20 sources
with the VLA. These twenty sources were selected based on 4.5 $\mu$m
Spitzer infrared images (see \S\ \ref{45data}). The angular resolution
of the 44 GHz VLA observations was 0.5$^{\prime\prime}$-1$^{\prime\prime}$,
the frequency resolution was 24 kHz ($\equiv$ 0.17 km s$^{-1}$),
and the rms noise was about 25 mJy beam$^{-1}$. The angular resolution
of the 6.7 GHz VLA observations was 2$^{\prime\prime}$ - 4$^{\prime\prime}$,
the frequency resolution was 3.05 kHz ($\equiv$0.14 km s$^{-1}$),
and the rms noise was about 27 mJy beam$^{-1}$. \citet{cyganowski2009}
found a strong association between the 4.5 $\mu$m selected objects
called EGOs (for ``Extended Green Objects'') and Class II masers,
and a widespread distribution of 44 GHz Class I masers in outflows,
indicating that EGOs are young, high mass star forming objects that
drive active outflows.

\subsection{The H$_{2}$O maser data}

Most of the H$_{2}$O maser data discussed in this paper were taken
from \citet{hoefner96} who obtained images and spectra for 21 H$_{2}$O
maser sources in the vicinity of UC H II regions by observing with
the VLA at 22 GHz. The angular resolution of their survey was 0.4$^{\prime\prime}$,
the frequency resolution was 24 kHz ($\equiv3.5$ km s$^{-1}$), and
the typical RMS noise was 30 mJy beam$^{-1}$.

\subsection{The UC H II region data}

Almost all of the UC H II region data in this paper were taken from
\citet{wood89}, who surveyed seventy-five UC H II regions with the
VLA at frequencies of 4.9 GHz and 14.9 GHz. The goal of this survey
was to understand the morphology and characteristics of the selected
H II regions. The 14.9 GHz observations used for this paper have
an angular resolution of approximately 0.4$^{\prime\prime}$ and an
rms noise of about 0.32 mJy beam$^{-1}$. In addition, the data for
one UC H II region in G45.47+0.07 were obtained from \citet{urquhart2009},
who observed 659 high mass star forming candidate regions with the
VLA at 4.9 GHz in order to detect UC H II regions. The angular resolution
of their observations was about 1.5$^{\prime\prime}$, and the rms
noise was about 0.2 mJy beam$^{-1}$. They identified 391 UC H II
regions in this sample.

\subsection{The 4.5 $\mu$m infrared data\label{45data}}

The 4.5 $\mu$m infrared data used in this paper were taken from
the GLIMPSE catalog of the online Spitzer archive (where GLIMPSE stands
for Galactic Legacy Infrared Midplane Survey Extraordinaire). The
GLIMPSE catalog (\citealt{churchwell2009}) was compiled from observations
taken with the Infrared Array Camera (IRAC) on board the Spitzer Space
Telescope, which recorded data at wavelenghts of 3.6 $\mu$m, 4.5
$\mu$m, 5.8 $\mu$m, and 8.0 $\mu$m. The angular resolution of GLIMPSE
images is about 1.2$^{\prime\prime}$. We use the 4.5 $\mu$m data
because \citet{cyganowski2008} have found that extended emission
at 4.5 $\mu$m is associated with high mass star forming regions.
This wavelength is usually colored green in most false color infrared
images, hence the term extended ``green'' objects (EGO's).

\section{Results}

\label{Chapter3}

We present the results. In \S\ \ref{statistical}, we discuss global
features of our data. In \S\ \ref{discussion}, we present a discussion
of eight individual regions for which high resolution data were available
for Class I and Class II methanol masers and associated tracers. In
\S\ \ref{lowres}, we present results for the regions for which
high angular resolution data were available only for Class I and Class
II methanol masers.

\subsection{Masers and associated objects. \label{statistical}}

\begin{figure}[h!]
\centering \includegraphics[width=0.9\textwidth]{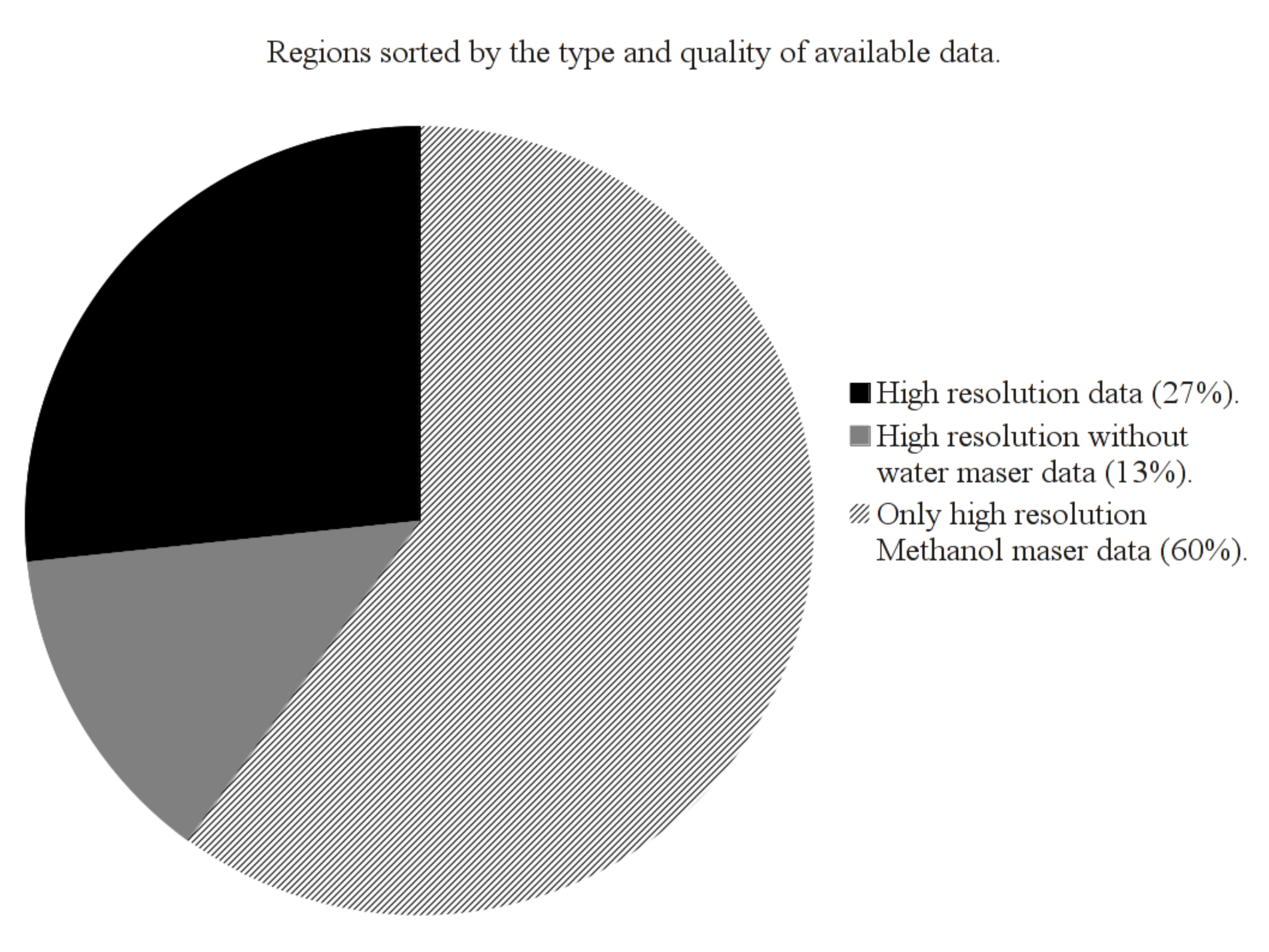}
\caption[Pie chart showing the type and quality of available data.]{Pie chart showing the type and quality of available data. In addition
to Class I and Class II methanol masers, we looked for data on H$_{2}$O
masers, UC H II regions, and 4.5 $\mu$m infrared sources. All of
these data were available for eight, or 27\%, of the regions. For
13\% of the regions, H$_{2}$O maser data were not found. For the
remaining 60\% of the regions, high resolution data were present for
the methanol masers only.}

\label{piechart} 
\end{figure}

The data for this project were obtained by searching the literature
for high angular resolution data on thirty star forming regions known
to contain Class I and Class II methanol masers. In addition to these
types of masers, we searched the literature for high angular resolution
data on associated H$_{2}$O masers, ultra compact (UC) ionized hydrogen
(H II) regions, and 4.5 $\mu$m Spitzer infrared images in these thirty
star forming regions. As shown in Figure \ref{piechart}, all of these
data were available for eight, or 27\% of the thirty regions. For
13\% of the regions, H$_{2}$O maser data, which were crucial to the
analyses we intended to perform, were not found, while data on methanol
masers, and UC H II regions were found. For the remaining 60\% of
the regions, high angular resolution data were present for the methanol
masers only.

\clearpage{}

\subsection{Discussion of individual regions.}

\label{discussion}

For those eight regions for which data are available in all the tracers
listed in \S\ \ref{statistical}, plots have been produced showing
the positions of the Class I and II methanol masers, H$_{2}$O masers,
UC H II regions and near infrared sources. Diagrams showing the center
velocities ($v_{\text{LSR}}$) and intensities of the methanol and
H$_{2}$O maser spectral lines were also produced for these regions.

With these plots of morphology and spectra we seek to address the
following issues:
\begin{enumerate}
\item What is a typical range of projected distances between Class I and
II methanol masers? 
\item How do the center velocities ($v_{\text{LSR}}$) of the H$_{2}$O
masers compare to that of the Class I and II methanol masers? 
\item How are H$_{2}$O masers arranged in relation to these methanol masers?
In particular, are there any arrangements that suggest a known configuration
like an outflow or disk? 
\end{enumerate}
The eight regions will now be discussed individually.

\subsection{Region G9.62+0.19}

\label{g9.62}

\begin{figure}[h!]
\centering \includegraphics[width=0.9\textwidth]{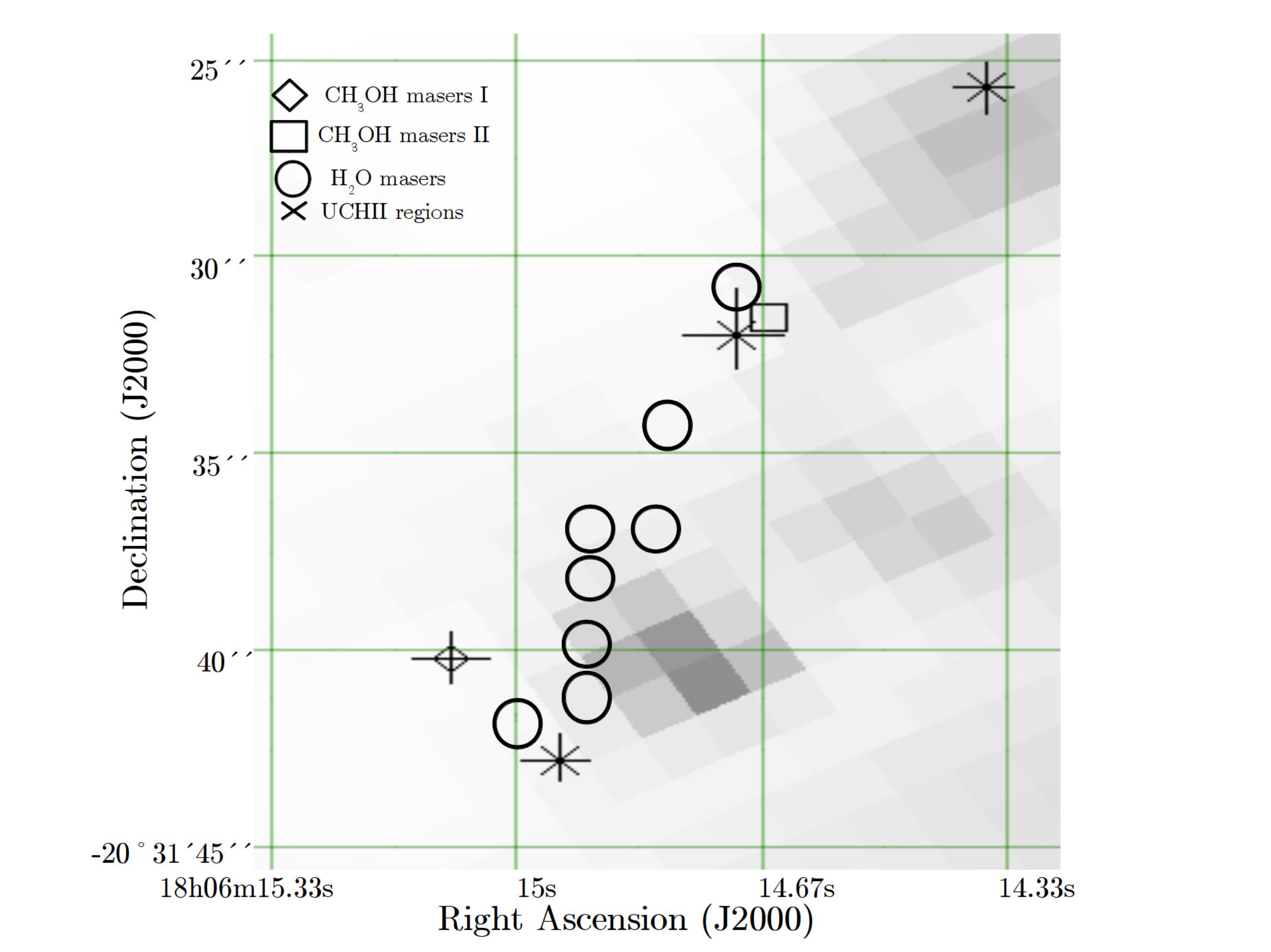}
\caption[Methanol masers and associated sources of interest in region G9.62+0.19]{Figure showing the Class II and the Class I methanol maser and other
sources of interest in region G9.62+0.19. The position of the Class
I methanol maser is indicated by a diamond and the Class II methanol
maser by a square. The open circles mark the positions of the H$_{2}$O
masers and the $\times$'s mark the peaks of the UC H II regions.
In all cases, positional uncertainties are either equal to or smaller
than the sizes of the markers themselves or are marked by a cross,
if larger than the marker sizes. The grayscale shows the 4.5 $\mu$m
Spitzer infrared image and the intensity range is between 1.61 MJy
sr$^{-1}$ and 938 MJy sr$^{-1}$. Per usual astronomical convention
for images, north is upward and east is to the left in this, and all
other images in this paper.}

 \label{f9.62} 
\end{figure}

Figure \ref{f9.62} shows the Class II methanol maser and the Class
I methanol maser in G9.62+0.19, together with other sources of interest.
To the northwest of the Class I maser position is a Class II methanol
maser, whose position is marked by a square in Figure \ref{f9.62}.
Following usual astronomical convention, east is to the left in Figure
\ref{f9.62}, and all other images in this paper. The Class I and
Class II masers are separated by about $10.7^{\prime\prime}$; at
a distance of 5.2 kiloparsec (kpc) to G9.62+0.19 (\citealt{sanna2009}),
this is equivalent to about 0.27 pc. The H$_{2}$O masers in this
source are arranged along a narrow, almost linear, structure (Figure
\ref{f9.62}). There are three UC H II regions in this figure (\citealt{forster2000}).
There is a UC H II region about $3.4^{\prime\prime}$ to the southwest
of the Class I methanol maser, and a second UC H II region that is
almost coincident with the Class II methanol maser. The grey scale
shows the 4.5 $\mu$m infrared Spitzer image. About $3.8^{\prime\prime}$
west of the Class I maser position lies a peak in this 4.5 $\mu$m
image. The positions of these sources, together with the telescopes
used to observe them and the angular resolution (in terms of synthesized
beam) of the observations, are listed in Table \ref{t9.62}.

\begin{figure}[h!]
\centering \includegraphics[width=0.9\textwidth]{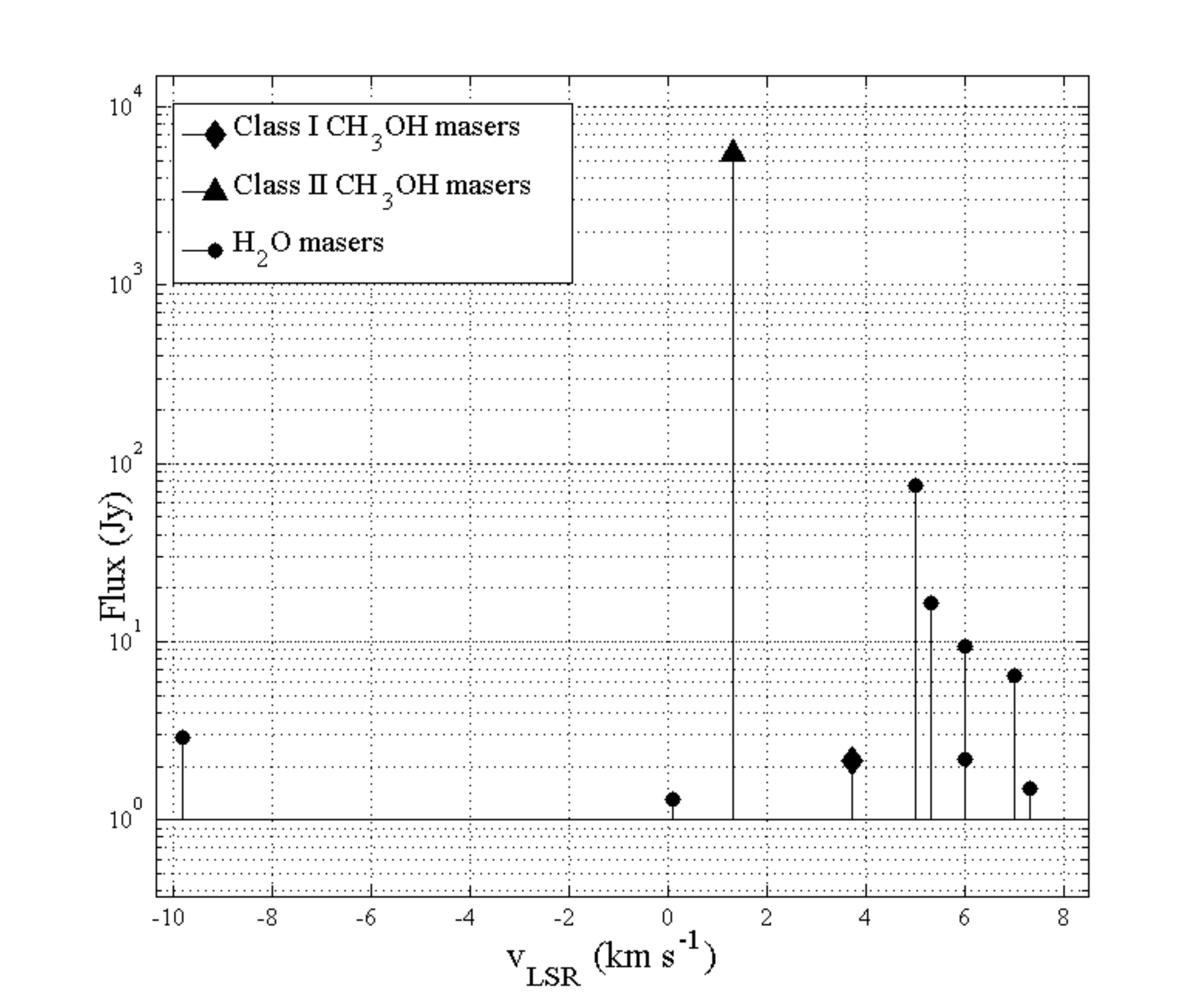}
\caption[Line profiles of methanol and H$_{2}$O masers in region G9.62+0.19]{Plot of intensities and center velocities ($v_{\text{LSR}}$) of
methanol and H$_{2}$O masers associated with G9.62+0.19. At the top
of each line is a symbol indicating the type of source; the Class
I methanol maser is represented by a diamond, the Class II methanol
maser by a triangle, and the H$_{2}$O masers by filled circles. A
H$_{2}$O maser with a center velocity of 25 km s$^{-1}$ is outside
the range of this plot. In this, and all other such figures, the $v_{\text{LSR}}$
of the Class II methanol maser will be used as a reference; masers
with $v_{\text{LSR}}$ larger than the Class II methanol maser will
be considered to be redshifted, and masers with $v_{\text{LSR}}$
smaller than the Class II methanol maser will be considered to be
blueshifted.}

\label{line9.62} 
\end{figure}

The center velocities ($v_{\text{LSR}}$) and intensities of the Class
I and Class II methanol masers and H$_{2}$O masers have been plotted
in Figure \ref{line9.62}. In this, and all other such figures, the
$v_{\text{LSR}}$ of the Class II methanol maser will be used as a
reference; masers with $v_{\text{LSR}}$ larger than the Class II
methanol maser will be considered to be redshifted, and masers with
$v_{\text{LSR}}$ smaller than the Class II methanol maser will be
considered to be blueshifted. Six of the eight H$_{2}$O masers in
G9.62+0.19 are at larger center velocities ($v_{\text{LSR}}$) than
the center velocity of the Class II methanol maser, that is, they
are redshifted with respect to the Class II methanol maser. The Class
I methanol maser is also redshifted (i.e., at larger $v_{\text{LSR}}$)
with respect to the Class II methanol maser. A ninth H$_{2}$O maser
with a center velocity of 25 km s$^{-1}$ is outside the range of
the plot in Figure \ref{line9.62}.

\providecommand{\tabularnewline}{\\}
\begin{deluxetable}{ccccccc}
\tablecolumns{7}
\tablewidth{0pc}
\tabletypesize{\small}
\rotate
\tablecaption{Positions and associated information for sources associated with G9.62+0.19\label{t9.62}}
\tablehead{
\colhead{Source Type}    &\colhead{R.A. (J2000)} &\colhead{Dec. (J2000)}&\colhead{Observing}&\colhead{Synthesized} & \multicolumn{2}{c}{Kinematic}\tabularnewline
\colhead{} & \colhead{(h m s)}   & \colhead{(${^{o}}~^{\prime}~^{\prime\prime}$)}    & \colhead{Telescope} &\colhead{Beam}    & \colhead{$v_{\text{LSR}}$ (km s$^{-1}$)} &\colhead{ Intensity (Jy)}}
\startdata

44 GHz Class I CH$_{3}$OH maser \tablenotemark{a} & 18 06 15.1  & $-$20 31 40  & VLA \tablenotemark{b} & $1.1^{\prime\prime}\times1.1^{\prime\prime}$  & 3.7  & 2.1\tabularnewline
12.2 GHz Class II CH$_{3}$OH maser \tablenotemark{c} & 18 06 14.7  & $-$20 31 32  & VLBA \tablenotemark{d}  & $0.001^{\prime\prime}\times0.001^{\prime\prime}$  & 1.3  & 5500\tabularnewline
22 MHz H$_{2}$O maser \tablenotemark{e} & 18 06 14.7  & $-$20 31 31  & VLA  & $0.64^{\prime\prime}\times0.33^{\prime\prime}$  & 0.10  & 1.3\tabularnewline
22 MHz H$_{2}$O maser  & 18 06 14.8  & $-$20 31 34  & VLA  & $0.64^{\prime\prime}\times0.33^{\prime\prime}$  & -9.8  & 2.9\tabularnewline
22 MHz H$_{2}$O maser  & 18 06 14.8  & $-$20 31 37  & VLA  & $0.64^{\prime\prime}\times0.33^{\prime\prime}$  & 5.3  & 17\tabularnewline
22 MHz H$_{2}$O maser  & 18 06 14.9  & $-$20 31 37  & VLA  & $0.64^{\prime\prime}\times0.33^{\prime\prime}$  & 6.0  & 2.2\tabularnewline
22 MHz H$_{2}$O maser  & 18 06 14.9  & $-$20 31 38  & VLA  & $0.64^{\prime\prime}\times0.33^{\prime\prime}$  & 7.3  & 1.5\tabularnewline
22 MHz H$_{2}$O maser  & 18 06 14.9  & $-$20 31 40  & VLA  & $0.64^{\prime\prime}\times0.33^{\prime\prime}$  & 7.0  & 6.4\tabularnewline
22 MHz H$_{2}$O maser  & 18 06 14.9  & $-$20 31 41  & VLA  & $0.64^{\prime\prime}\times0.33^{\prime\prime}$  & 5.0  & 75 \tabularnewline
22 MHz H$_{2}$O maser  & 18 06 14.9  & $-$20 31 41  & VLA  & $0.64^{\prime\prime}\times0.33^{\prime\prime}$  & 6.0  & 9.5\tabularnewline
22 MHz H$_{2}$O maser  & 18 06 15.0  & $-$20 31 42  & VLA  & $0.64^{\prime\prime}\times0.33^{\prime\prime}$  & 25  & 0.30\tabularnewline
8.2 GHz and 9.2 GHz UCHII region \tablenotemark{f} & 18 06 14.7  & $-$20 31 32  & ATCA \tablenotemark{g} & $1.1^{\prime\prime}\times1.2^{\prime\prime}$  & \nodata\tablenotemark{h} & \nodata\tablenotemark{h}\tabularnewline
8.2 GHz and 9.2 GHz UCHII region  & 18 06 14.4  & $-$20 31 26  & ATCA  & $2.2^{\prime\prime}\times1.6^{\prime\prime}$  & \nodata  & \nodata\tabularnewline
8.2 GHz and 9.2 GHz UCHII region  & 18 06 14.9  & $-$20 31 43  & ATCA  & $0.7^{\prime\prime}\times0.6^{\prime\prime}$  & \nodata  & \nodata\tabularnewline
 &  &  &  &  &  & \tabularnewline
\enddata
\par

\tablenotetext{a}{All 44GHz class I methanol masers in this section were found
in \citet{valtts2007} and \citet{kurtz2004}}

\tablenotetext{b}{The Karl G. Jansky Very Large Array.}

\tablenotetext{c}{\citet{minier2001}}

\tablenotetext{d}{The Very Long Baseline Array.}

\tablenotetext{e}{Information on all water masers taken from \citet{hoefner96}
unless otherwise noted. }

\tablenotetext{f}{All of these UC H II regions were reported in \citet{forster2000}.
\citet{forster2000} explain that their observations were taken at
both 8.2 and 9.2 GHz.}

\tablenotetext{g}{The Australia Telescope Compact Array (ATCA).}

\tablenotetext{h}{UC HII regions are continuum data and, therefore, produce no
line profile information.}

\tablecomments{Sources consulted for data presented in tables one through eight reported different
significant figures. For convenience the data have been rounded to
the least number of significant figures. Information on the resolution
of each observing instrument is presented in the column `synthesized
beam' which differs between instruments and configurations of the
same instrument.}
\end{deluxetable}


\subsection{Region G10.47+0.03}

\label{g10.47}

\begin{figure}[h!]
\centering \includegraphics[width=0.9\textwidth]{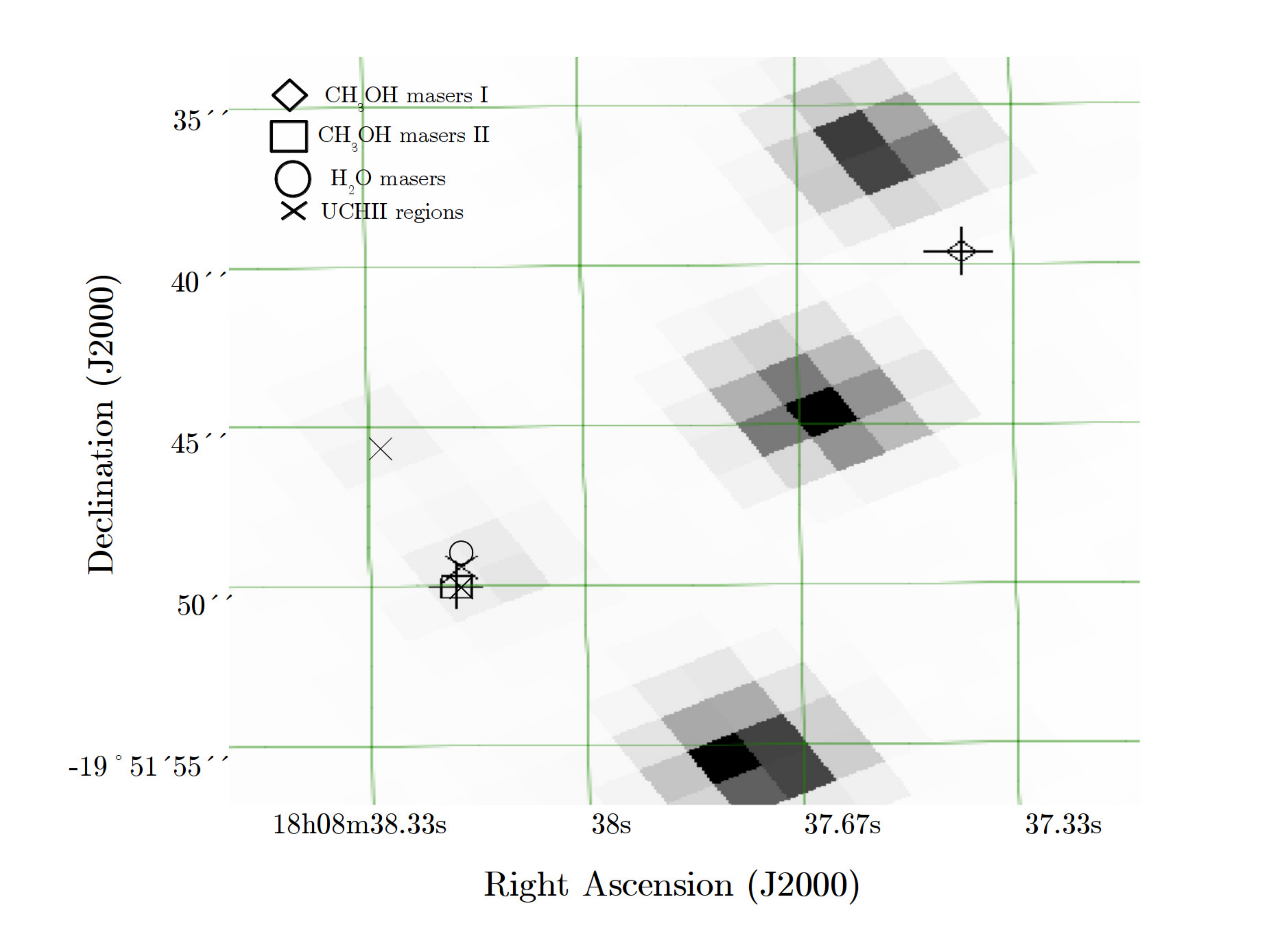}
\caption[Methanol masers and associated sources of interest in region G10.47+0.03]{Figure showing the Class II and the Class I methanol maser and other
sources of interest in region G10.47+0.03. The symbols are the same
as in Figure \ref{f9.62}, and are also marked in a box at the top
left of the figure. The grayscale shows the 4.5 $\mu$m Spitzer infrared
image and the intensity range is between 2.22 MJy sr$^{-1}$ and 1100
MJy sr$^{-1}$.}

 \label{f10.47} 
\end{figure}

Figure \ref{f10.47} shows the Class II methanol maser and the Class
I methanol maser in G10.47+0.03, together with other sources of interest.
The Class II methanol maser is located to the southeast of the Class
I maser position, and its position is marked by a square in Figure
\ref{f10.47}. The Class I and Class II masers are separated by about
$15.3^{\prime\prime}$; at a distance of 6.0 kpc to G10.47+0.03 (\citealt{pestalozzi2005}),
this is equivalent to about 0.44 pc. Two of the H$_{2}$O masers in
this source are almost coincident with the Class II methanol maser,
and a third is right next to it. There are three UC H~II regions,
two of which are almost coincident with the Class II methanol maser,
and a third lies to the northeast of it. A line of strong 4.5 $\mu$m
point sources lies to the east and southeast of the Class I methanol
maser. The positions of these sources, together with the telescopes
used to observe them and the angular resolution in terms of synthesized
beam of the observations, are listed in Table \ref{t10.47}.

\begin{figure}[h!]
\centering \includegraphics[width=0.9\textwidth]{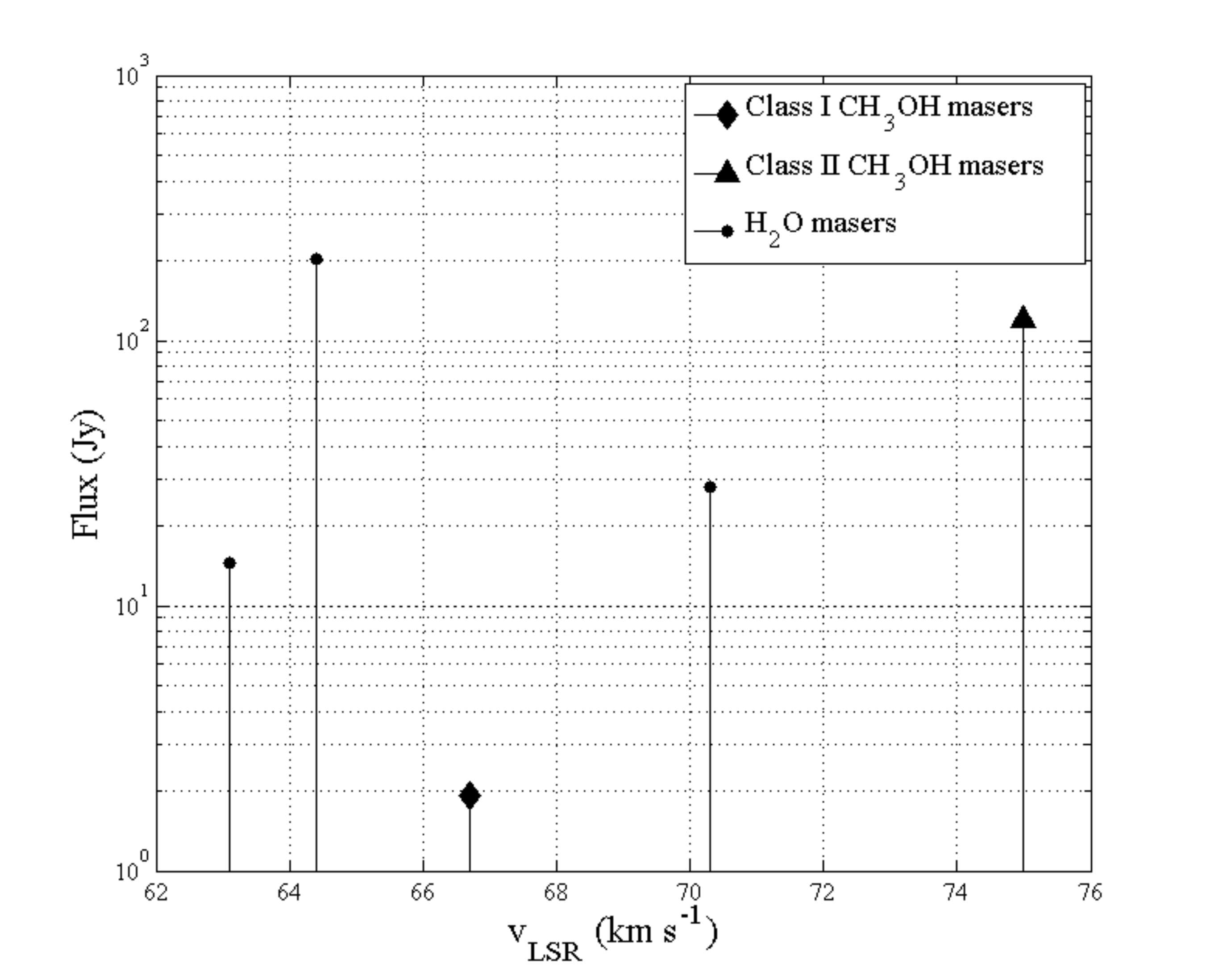}
\caption[Line profiles of methanol and H$_{2}$O masers in region G10.47+0.03]{Plot of intensities and center velocities ($v_{\text{LSR}}$) of
methanol and H$_{2}$O masers associated with G10.47+0.03. The symbols
are the same as in Figure \ref{line9.62}.}

\label{line10.47} 
\end{figure}

The center velocities ($v_{\text{LSR}}$) and intensities of the Class
I and Class II methanol masers and H$_{2}$O masers have been plotted
in Figure \ref{line10.47}. All the three H$_{2}$O masers shown in
this figure are at smaller $v_{\text{LSR}}$ (i.e., blueshifted) than
the Class II methanol maser. The Class I methanol maser is also blueshifted
with respect to the Class II methanol maser.

\notetoeditor{In a previous version the label on this region was a typo.  It was G10.47+0.27 when I clearly meant G10.47+0.03}

\providecommand{\tabularnewline}{\\}
\begin{deluxetable}{ccccccc}
\tablecolumns{7}
\tablewidth{0pc}
\tabletypesize{\small}
\rotate
\tablecaption{Positions and associated information for sources associated with G10.47+0.03\label{t10.47}}
\tablehead{
\colhead{Source Type}    &\colhead{R.A. (J2000)} &\colhead{Dec. (J2000)}&\colhead{Observing}&\colhead{Synthesized} & \multicolumn{2}{c}{Kinematic}\tabularnewline
\colhead{} & \colhead{(h m s)}   & \colhead{(${^{o}}~^{\prime}~^{\prime\prime}$)}    & \colhead{Telescope} &\colhead{Beam}    & \colhead{$v_{\text{LSR}}$ (km s$^{-1}$)} &\colhead{ Intensity (Jy)}}
\startdata

44 GHz Class I CH$_{3}$OH maser \tablenotemark{a} & 18 08 37.4  & -19 51 40  & VLA  & $3.14^{\prime\prime}\times1.33^{\prime\prime}$  & 67 & 1.9\tabularnewline
6.7 GHz Class II CH$_{3}$OH maser \tablenotemark{b}  & 18 08 38.2  & -19 51 50  & ATCA  & $1.5^{\prime\prime}\times1.5^{\prime\prime}$  & 75 & 120\tabularnewline
22 GHz H$_{2}$O maser \tablenotemark{c} & 18 08 38.2  & -19 51 50  & VLA  & $0.48^{\prime\prime}\times0.45^{\prime\prime}$  & 64 & 200\tabularnewline
22 GHz H$_{2}$O maser  & 18 08 38.2  & -19 51 50  & VLA  & $0.48^{\prime\prime}\times0.45^{\prime\prime}$  & 63 & 14\tabularnewline
22 GHz H$_{2}$O maser  & 18 08 38.2  & -19 51 49  & VLA  & $0.48^{\prime\prime}\times0.45^{\prime\prime}$  & 70 & 28\tabularnewline
14.9 GHz UCHII region \tablenotemark{d} & 18 08 38.2  & -19 51 50 & VLA  & $0.89^{\prime\prime}\times0.57^{\prime\prime}$  & \nodata& \nodata\tabularnewline
14.9 GHz UCHII region  & 18 08 38.2  & -19 51 49 & VLA  & $0.79^{\prime\prime}\times0.56^{\prime\prime}$  & \nodata & \nodata \tabularnewline
14.9 GHz UCHII region  & 18 08 38.3  & -19 51 45 & VLA  & $1.04^{\prime\prime}\times0.79^{\prime\prime}$  & \nodata & \nodata\tabularnewline
 &  &  &  &  &  & \tabularnewline
\enddata
\par

\tablenotetext{a}{\citet{kurtz2004} } 
\tablenotetext{b}{\citet{caswell952} }
\tablenotetext{c}{\citet{hoefner96} }
\tablenotetext{d}{\citet{wood89} }
\end{deluxetable}

\subsection{Region G12.20-0.11}

\label{g12.20}

\begin{figure}[h!]
\centering \includegraphics[width=0.9\textwidth]{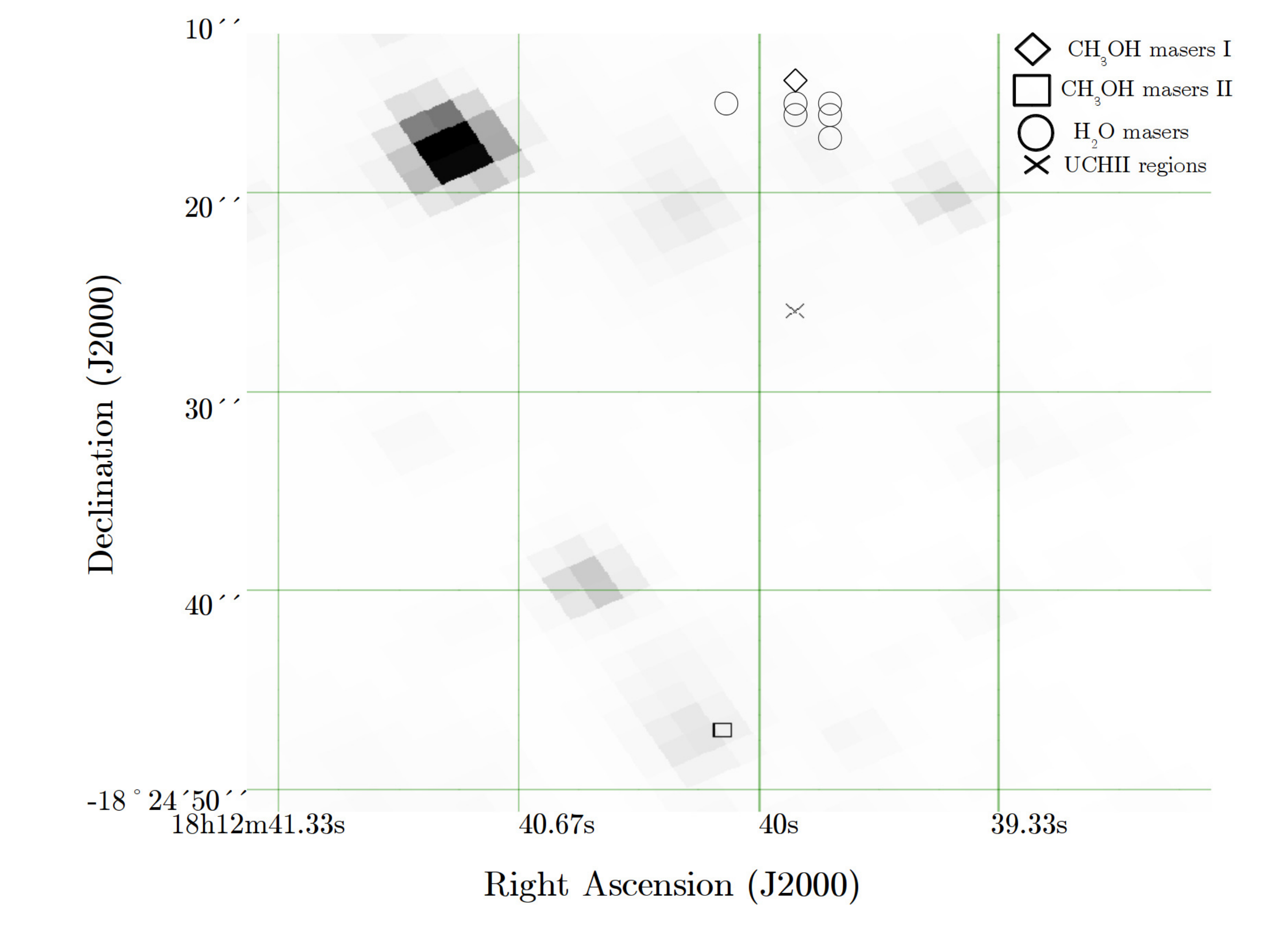}
\caption[Methanol masers and associated sources of interest in region G12.20-0.11]{Figure showing the Class II and the Class I methanol maser and other
sources of interest in region G12.20-0.11. The grayscale shows the
4.5 $\mu$m Spitzer infrared image and the intensity range is between
2.58 MJy sr$^{-1}$ and 1220 MJy sr$^{-1}$. The symbols are the same
as in Figure \ref{f9.62}, and are also marked in a box at the top
right of the figure.}

 \label{f12.20} 
\end{figure}

\begin{figure}[h!]
\centering \includegraphics[width=0.9\textwidth]{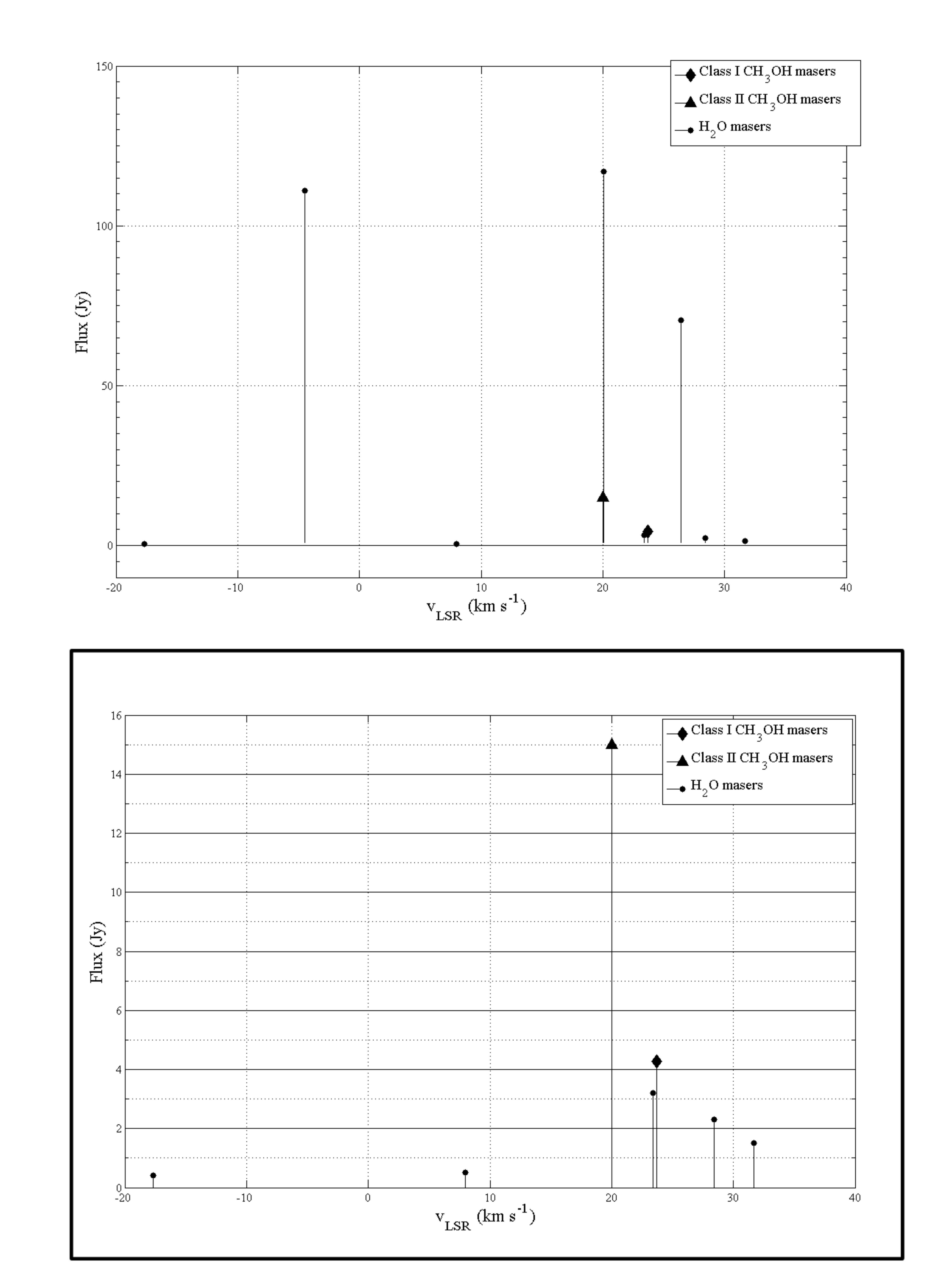}
\caption[Line profiles of methanol and H$_{2}$O masers in region G12.20-0.11]{Plot of intensities and center velocities ($v_{\text{LSR}}$) of
methanol and H$_{2}$O masers associated with G12.20-0.11. The symbols
are the same as in Figure \ref{line9.62}. For better visibility,
an enlarged version of the 0-16 Jy segment in the upper plot is presented
in the lower plot.}

\label{line12.20} 
\end{figure}

Figure \ref{f12.20} shows the Class II methanol maser and the Class
I methanol maser in G12.20-0.11, together with other sources of interest.
The Class I and Class II masers are separated by about $32.4^{\prime\prime}$;
at a distance of 8.3 kpc to G12.20-0.11 (\citealt{pestalozzi2005}),
this is equivalent to about 1.3 pc. Six of the H$_{2}$O masers in
this source are clustered around the Class I methanol maser. An ultracompact
HII region  is to the south of the Class I methanol
maser. To the east of the Class I methanol maser there is a strong
4.5 $\mu$m infrared source. Figure \ref{line12.20} shows the line
velocities of the masers in this region. The positions of these sources,
together with the telescopes used to observe them and the angular
resolution (in terms of synthesized beam) of the observations, are
listed in Table \ref{t12.20}.

The center velocities ($v_{\text{LSR}}$) and intensities of the Class
I and Class II methanol masers and H$_{2}$O masers have been plotted
in Figure \ref{line12.20}. Four of the eight H$_{2}$O masers are
at $v_{\text{LSR}}$ greater than (i.e., redshifted) the $v_{\text{LSR}}$
of the Class II methanol maser. The Class I methanol maser is also
redshifted with respect to the Class II methanol maser.

\providecommand{\tabularnewline}{\\}
\begin{deluxetable}{ccccccc}
\tablecolumns{7}
\tablewidth{0pc}
\tabletypesize{\small}
\rotate
\tablecaption{Positions and associated information for sources associated with G12.20-0.11\label{t12.20}}
\tablehead{
\colhead{Source Type}    &\colhead{R.A. (J2000)} &\colhead{Dec. (J2000)}&\colhead{Observing}&\colhead{Synthesized} & \multicolumn{2}{c}{Kinematic}\tabularnewline
\colhead{} & \colhead{(h m s)}   & \colhead{(${^{o}}~^{\prime}~^{\prime\prime}$)}    & \colhead{Telescope} &\colhead{Beam}    & \colhead{$v_{\text{LSR}}$ (km s$^{-1}$)} &\colhead{ Intensity (Jy)}}
\startdata
44 GHz Class I CH$_{3}$OH maser \tablenotemark{a} & 18 12 39.9  & -18 24 15  & VLA  & $2.93^{\prime\prime}\times1.43^{\prime\prime}$  & 24 & 4.3\tabularnewline
6.7 GHz Class II CH$_{3}$OH maser \tablenotemark{b}  & 18 12 40.1  & -18 24 47  & ATCA  & $1.5^{\prime\prime}\times1.5^{\prime\prime}$  & 20 & 15\tabularnewline
22 GHz H$_{2}$O maser \tablenotemark{c} & 18 12 39.9  & -18 24 17  & VLA  & $0.49^{\prime\prime}\times0.38^{\prime\prime}$  & 20 & 120\tabularnewline
22 GHz H$_{2}$O maser  & 18 12 39.8  & -18 24 17  & VLA  & $0.49^{\prime\prime}\times0.38^{\prime\prime}$  & -4.5 & 110\tabularnewline
22 GHz H$_{2}$O maser  & 18 12 44.5  & -18 24 25  & VLA  & $0.49^{\prime\prime}\times0.38^{\prime\prime}$  & 26 & 70\tabularnewline
22 GHz H$_{2}$O maser  & 18 12 39.8  & -18 24 18  & VLA  & $0.49^{\prime\prime}\times0.38^{\prime\prime}$  & 23 & 3.2\tabularnewline
22 GHz H$_{2}$O maser  & 18 12 39.8  & -18 24 16  & VLA  & $0.49^{\prime\prime}\times0.38^{\prime\prime}$  & 32 & 1.5\tabularnewline
22 GHz H$_{2}$O maser  & 18 12 44.5  & -18 24 24  & VLA  & $0.49^{\prime\prime}\times0.38^{\prime\prime}$  & 28 & 2.3\tabularnewline
22 GHz H$_{2}$O maser  & 18 12 39.9  & -18 24 16  & VLA  & $0.49^{\prime\prime}\times0.38^{\prime\prime}$  & 8.0 & 0.50\tabularnewline
22 GHz H$_{2}$O maser  & 18 12 40.1  & -18 24 16  & VLA  & $0.49^{\prime\prime}\times0.38^{\prime\prime}$  & -18 & 0.40\tabularnewline
14.9 GHz UCHII region \tablenotemark{d} & 18 12 38.1  & -18 25 10  & VLA  & $0.4^{\prime\prime}\times0.4^{\prime\prime}$  &\nodata & \nodata\tabularnewline
14.9 GHz UCHII region  & 18 12 43.7  & -18 25 44  & VLA  & $0.4^{\prime\prime}\times0.4^{\prime\prime}$  & \nodata & \nodata \tabularnewline
14.9 GHz UCHII region  & 18 12 39.9  & -18 24 26  & VLA  & $0.4^{\prime\prime}\times0.4^{\prime\prime}$  & \nodata & \nodata\tabularnewline
 &  &  &  &  &  & \tabularnewline
\enddata
\par
\tablenotetext{a}{\citet{kurtz2004} } 
\tablenotetext{b}{\citet{caswell952} }
\tablenotetext{c}{\citet{hoefner96} }
\tablenotetext{d}{\citet{wood89} }
\end{deluxetable}

\subsection{Region G31.41+0.31}

\label{g31.41}

\begin{figure}[h!]
\centering \includegraphics[width=0.9\textwidth]{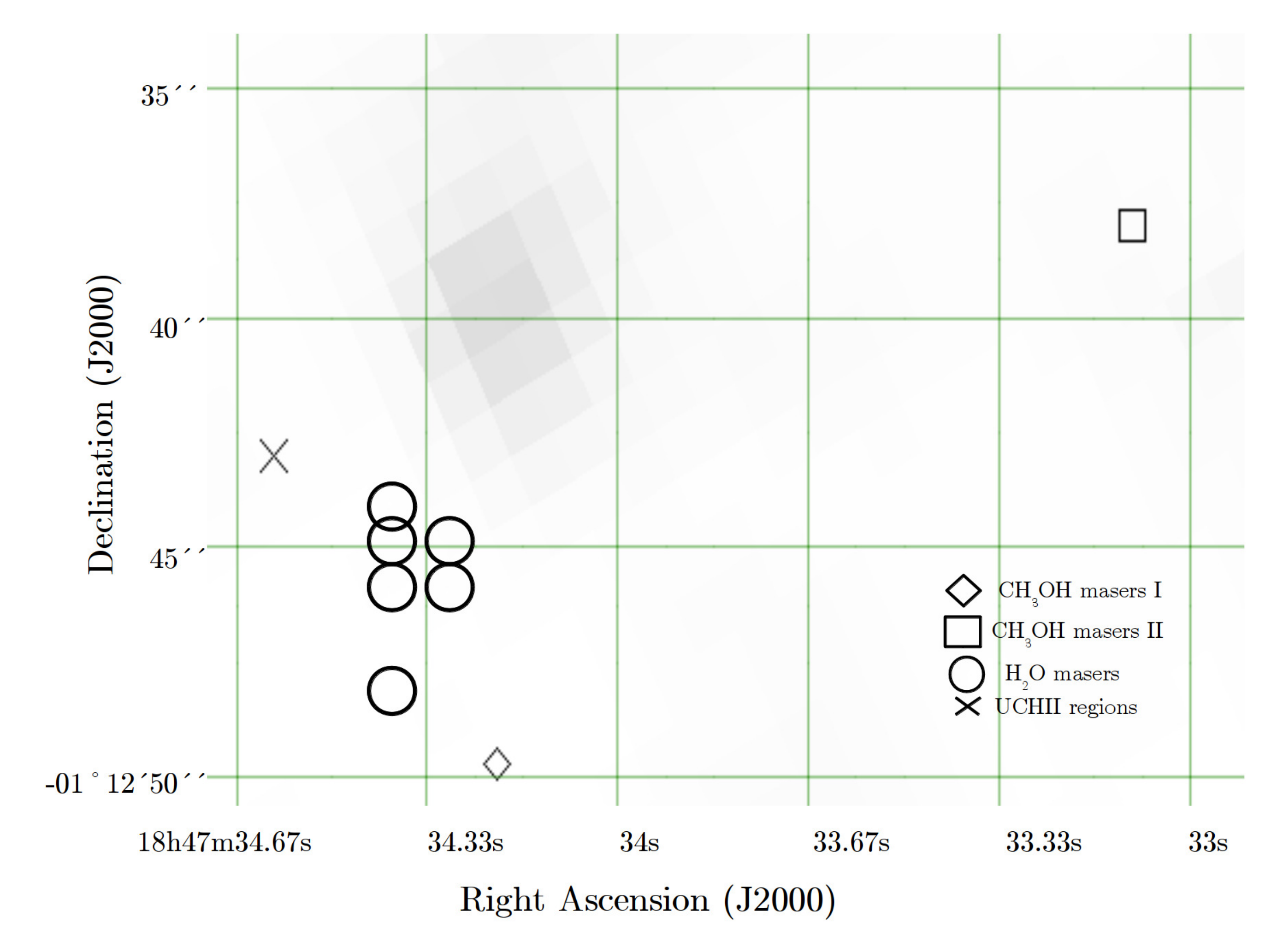}
\caption[Methanol masers and associated sources of interest in region G31.41+0.31]{Figure showing the Class II and the Class I methanol maser and other
sources of interest in region G31.41+0.31. The grayscale shows the
4.5 $\mu$m Spitzer infrared image and the intensity range is between
1.13 MJy sr$^{-1}$ and 658 MJy sr$^{-1}$. The symbols are the same
as in Figure \ref{f9.62}.}

 \label{f31.41} 
\end{figure}

Figure \ref{f31.41} shows the Class II methanol maser and the Class
I methanol maser in G31.41+0.31, together with other sources of interest.
The Class I methanol maser lies about $20.43^{\prime\prime}$ to the
southeast of the Class I methanol maser; at a distance of 7.3 kpc
to G31.41+0.31 (\citealt{pestalozzi2005}), this is equivalent to
about 0.72 pc. A complex of H$_{2}$O masers is about $5^{\prime\prime}$
north of the Class I methanol maser position. A UC HII region is $7^{\prime\prime}$
to the northeast of the Class I methanol maser. This region shows
two weak 4.5 $\mu$m infrared sources, one about 10$^{\prime\prime}$
north of the Class I methanol maser and the other to the west of the
Class II methanol maser. The positions of these sources, together
with the telescopes used to observe them and the angular resolution
(in terms of synthesized beam) of the observations, are listed in
Table \ref{t31.41}.

\begin{figure}[h!]
\centering \includegraphics[width=0.9\textwidth]{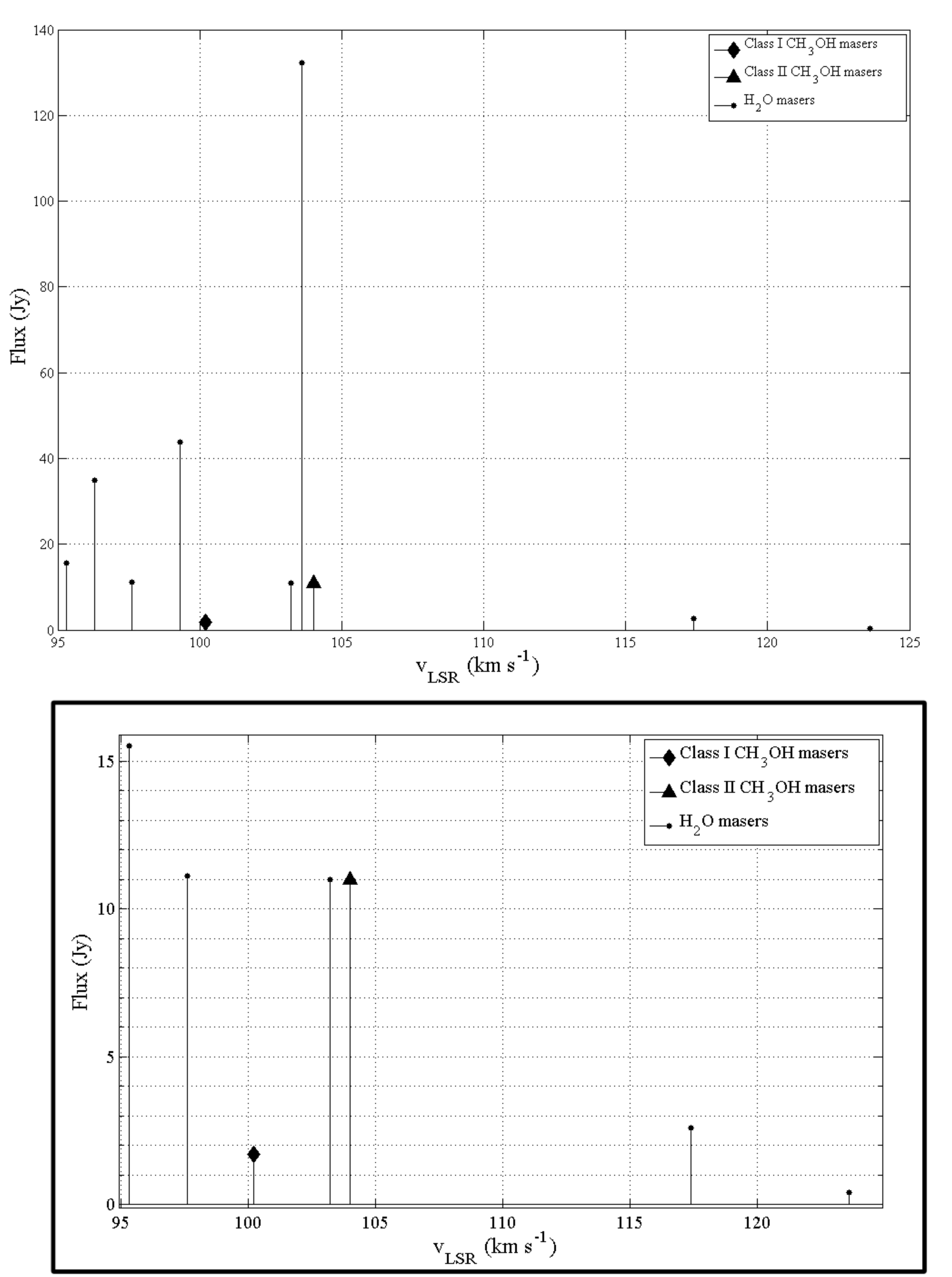}
\caption[Line profiles of methanol and H$_{2}$O masers in region G31.41+0.31]{Plot of intensities and center velocities ($v_{\text{LSR}}$) of
methanol and H$_{2}$O masers associated with G31.41+0.31. The symbols
are the same as in Figure \ref{line9.62}. The lower plot is an enlarged
version of the 0-15 Jy segment in the upper plot.}

\label{line31.41} 
\end{figure}

The center velocities ($v_{\text{LSR}}$) and intensities of the Class
I and Class II methanol masers and H$_{2}$O masers have been plotted
in Figure \ref{line31.41}. Six of the eight H$_{2}$O masers shown
in this figure are at lower $v_{\text{LSR}}$ (i.e., blueshifted)
than the Class II methanol maser. The Class I methanol maser is also
blueshifted with respect to the Class II methanol maser.

\providecommand{\tabularnewline}{\\}
\begin{deluxetable}{ccccccc}
\tablecolumns{7}
\tablewidth{0pc}
\tabletypesize{\small}
\rotate
\tablecaption{Positions and associated information for sources associated with G31.41+0.31\label{t31.41}}
\tablehead{
\colhead{Source Type}    &\colhead{R.A. (J2000)} &\colhead{Dec. (J2000)}&\colhead{Observing}&\colhead{Synthesized} & \multicolumn{2}{c}{Kinematic}\tabularnewline
\colhead{} & \colhead{(h m s)}   & \colhead{(${^{o}}~^{\prime}~^{\prime\prime}$)}    & \colhead{Telescope} &\colhead{Beam}    & \colhead{$v_{\text{LSR}}$ (km s$^{-1}$)} &\colhead{ Intensity (Jy)}}
\startdata
44 GHz Class I CH$_{3}$OH maser \tablenotemark{a} & 18 47 34.2 & -01 12 50  & VLA  & $2.21^{\prime\prime}\times1.36^{\prime\prime}$  & 100 & 1.7\tabularnewline
6.7 GHz Class II CH$_{3}$OH maser \tablenotemark{b}  & 18 47 33.1  & -01 12 38  & ATCA  & $1.5^{\prime\prime}\times1.5^{\prime\prime}$  & 104  & 11\tabularnewline
22 GHz H$_{2}$O maser \tablenotemark{c} & 18 47 34.3  & -01 12 45  & VLA  & $0.47^{\prime\prime}\times0.47^{\prime\prime}$  & 104  & 130\tabularnewline
22 GHz H$_{2}$O maser  & 18 47 34.3  & -01 12 46  & VLA  & $0.47^{\prime\prime}\times0.47^{\prime\prime}$  & 103  & 11\tabularnewline
22 GHz H$_{2}$O maser  & 18 47 34.3  & -01 12 46  & VLA  & $0.47^{\prime\prime}\times0.47^{\prime\prime}$  & 95.3  & 16\tabularnewline
22 GHz H$_{2}$O maser  & 18 47 34.3  & -01 12 46  & VLA  & $0.47^{\prime\prime}\times0.47^{\prime\prime}$  & 96.3  & 35\tabularnewline
22 GHz H$_{2}$O maser  & 18 47 34.4  & -01 12 44  & VLA  & $0.47^{\prime\prime}\times0.47^{\prime\prime}$  & 117  & 2.6\tabularnewline
22 GHz H$_{2}$O maser  & 18 47 34.4  & -01 12 45  & VLA  & $0.47^{\prime\prime}\times0.47^{\prime\prime}$  & 99.3  & 44\tabularnewline
22 GHz H$_{2}$O maser  & 18 47 34.4  & -01 12 46  & VLA  & $0.47^{\prime\prime}\times0.47^{\prime\prime}$  & 97.6  & 11\tabularnewline
22 GHz H$_{2}$O maser  & 18 47 34.4  & -01 12 48  & VLA  & $0.47^{\prime\prime}\times0.47^{\prime\prime}$  & 124  & 0.40\tabularnewline
14.9 GHz UCHII region \tablenotemark{d} & 18 47 34.6  & -01 12 43  & VLA  & $1.1^{\prime\prime}\times1.1^{\prime\prime}$  & \nodata & \nodata\tabularnewline
 &  &  &  &  &  & \tabularnewline
\enddata
\par
\tablenotetext{a}{\citet{kurtz2004} } 
\tablenotetext{b}{\citet{caswell952} }
\tablenotetext{c}{\citet{hoefner96} }
\tablenotetext{d}{\citet{wood89} }
\end{deluxetable}

\subsection{Region G35.03+0.35}

\label{g35.03}

\begin{figure}[h!]
\centering \includegraphics[width=0.9\textwidth]{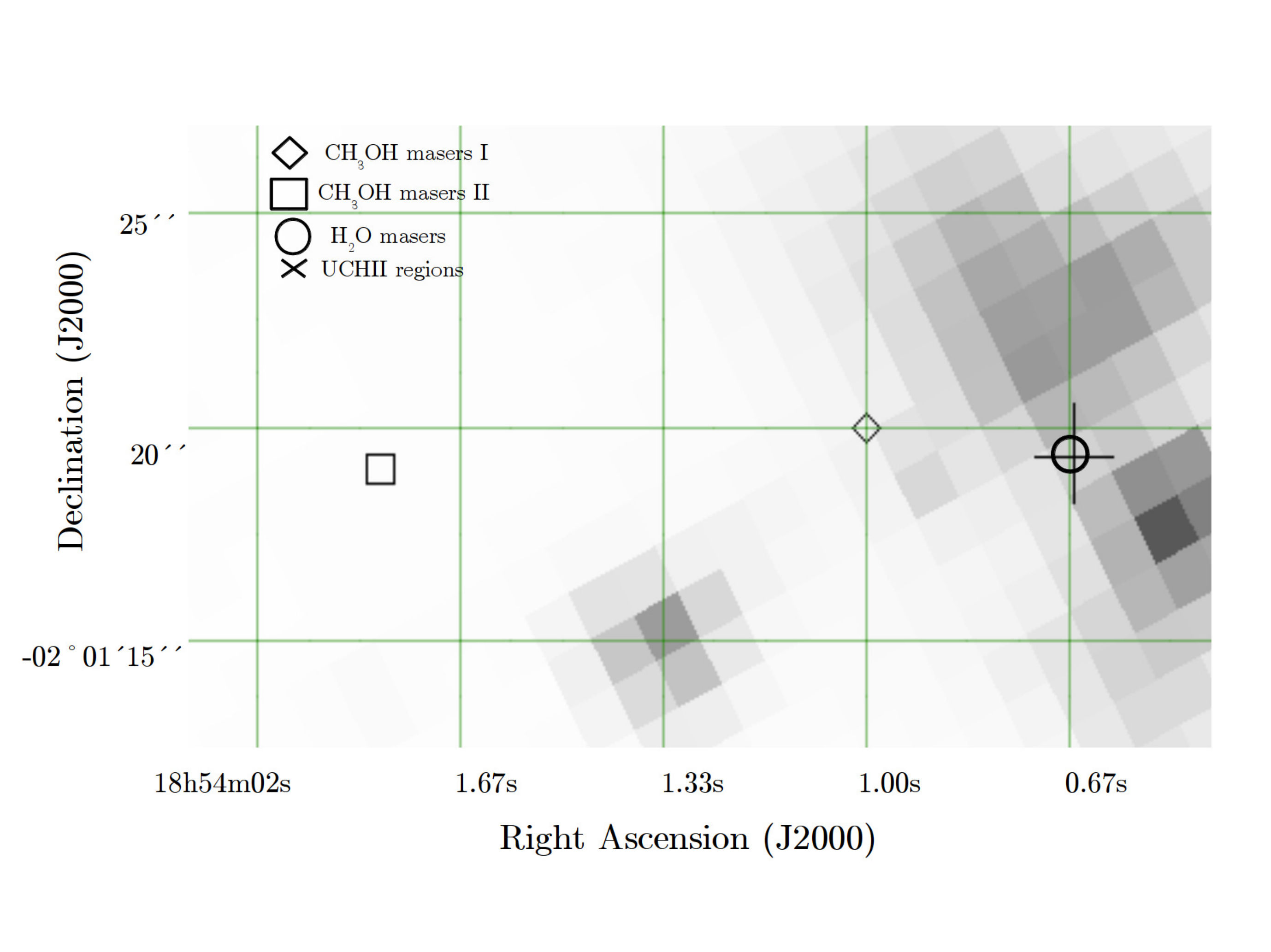}
\caption[Methanol masers and associated sources of interest in region G35.03+0.35]{Figure showing the Class II and the Class I methanol maser and other
sources of interest in region G35.03+0.35. The grayscale shows the
4.5 $\mu$m Spitzer infrared image and the intensity range is between
0.374 MJy sr$^{-1}$ and 455 MJy sr$^{-1}$. The symbols are the same
as in Figure \ref{f9.62}.}

 \label{f35.03} 
\end{figure}

Figure \ref{f35.03} shows the Class II methanol maser and the Class
I methanol maser in G35.03+0.35, together with other sources of interest.
The Class I methanol maser lies almost due west of the Class II methanol
maser by about $11.98^{\prime\prime}$; at a distance of 6.9 kpc to
G35.03+0.35 (\citealt{pestalozzi2005}), this is equivalent to about
0.40 pc. There is only one H$_{2}$O maser in this source, and it
lies to the west of the Class I methanol maser. Extended 4.5~$\mu$m
emission straddles the region between the Class I methanol and H$_{2}$O
maser. The positions of these sources, together with the telescopes
used to observe them and the angular resolution (in terms of synthesized
beam) of the observations, are listed in Table \ref{t35.03}.

\begin{figure}[h!]
\centering \includegraphics[width=0.9\textwidth]{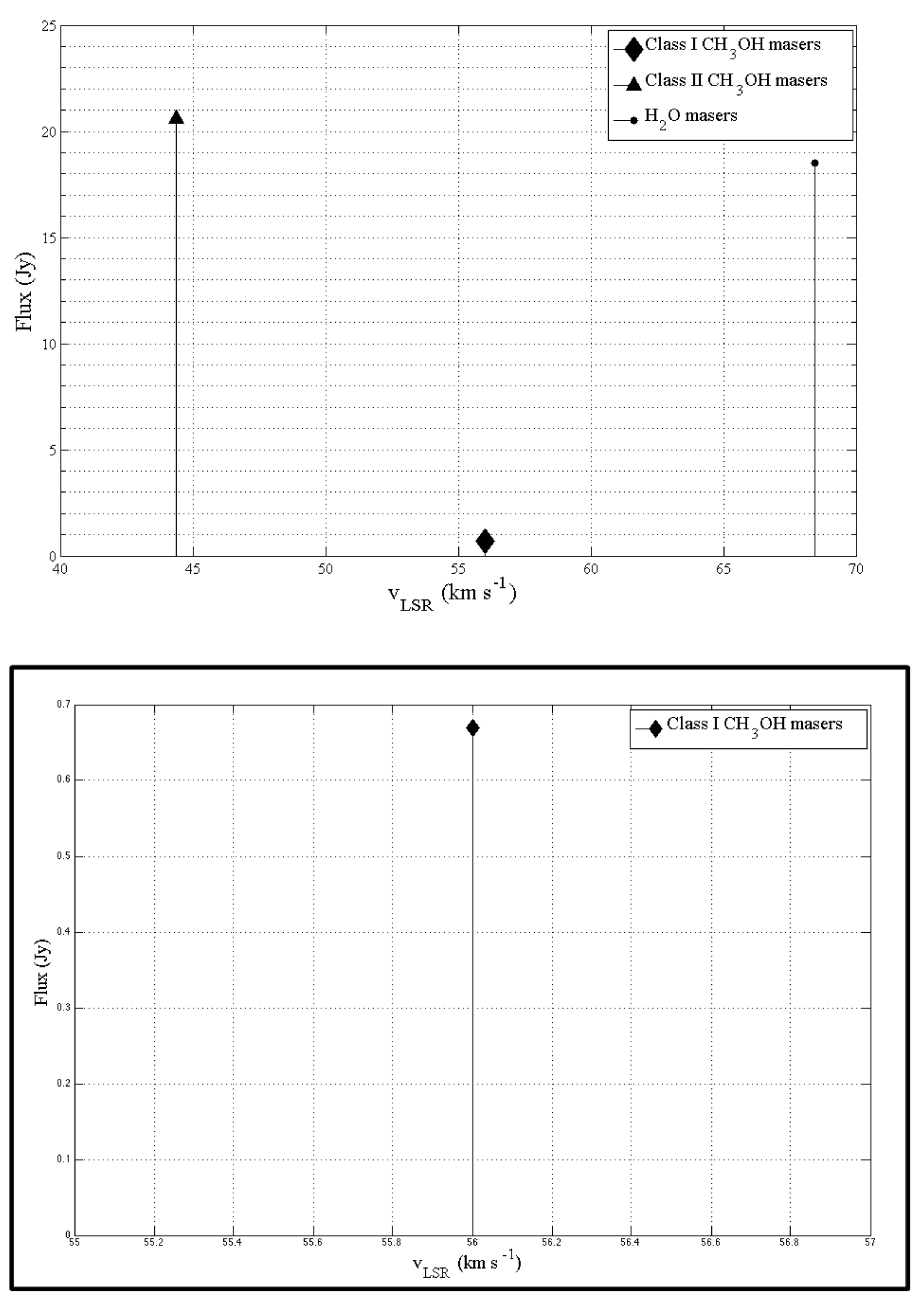}
\caption[Line profiles of methanol and H$_{2}$O masers in region G35.03+0.35]{Plot of intensities and center velocities ($v_{\text{LSR}}$) of
methanol and H$_{2}$O masers associated with G35.03+0.35. The symbols
are the same as in Figure \ref{line9.62}. The lower plot is an enlarged
version of the 0-0.7 Jy segment in the upper plot.}

\label{line35.03} 
\end{figure}

The center velocities ($v_{\text{LSR}}$) and intensities of the Class
I and Class II methanol masers and H$_{2}$O masers have been plotted
in Figure \ref{line35.03}. The H$_{2}$O maser is at a higher $v_{\text{LSR}}$
(i.e., redshifted) than the Class II methanol maser. The Class I methanol
maser is also redshifted with respect to the Class II methanol maser.

\providecommand{\tabularnewline}{\\}
\begin{deluxetable}{ccccccc}
\tablecolumns{7}
\tablewidth{0pc}
\tabletypesize{\small}
\rotate
\tablecaption{Positions and associated information for sources associated with G35.03+0.35\label{t35.03} }
\tablehead{
\colhead{Source Type}    &\colhead{R.A. (J2000)} &\colhead{Dec. (J2000)}&\colhead{Observing}&\colhead{Synthesized} & \multicolumn{2}{c}{Kinematic}\tabularnewline
\colhead{} & \colhead{(h m s)}   & \colhead{(${^{o}}~^{\prime}~^{\prime\prime}$)}    & \colhead{Telescope} &\colhead{Beam}    & \colhead{$v_{\text{LSR}}$ (km s$^{-1}$)} &\colhead{ Intensity (Jy)}}
\startdata

44 GHz Class I CH$_{3}$OH maser \tablenotemark{a}  & 18 54 1.0  & +02 01 20  & VLA  & $0.58^{\prime\prime}\times0.46^{\prime\prime}$  & 56  & 0.67\tabularnewline
6.7 GHz Class II CH$_{3}$OH maser \tablenotemark{b}   & 18 54 1.8  & +02 01 19  & VLA  & $2.32^{\prime\prime}\times1.34^{\prime\prime}$  & 44  & 21\tabularnewline
22 GHz H$_{2}$O maser \tablenotemark{c}  & 18 54 0.66  & +02 01 19  & VLA  & $2.6^{\prime\prime}\times1.8^{\prime\prime}$  & 68 & 19\tabularnewline
14.9 GHz UCHII region \tablenotemark{d}  & 18 51 29  & +01 57 31  & VLA  & $0.93^{\prime\prime}\times0.8^{\prime\prime}$  & \nodata  & \nodata\tabularnewline
 &  &  &  &  &  & \tabularnewline
\enddata
\par
\tablenotetext{a}{\citet{kurtz2004} } 
\tablenotetext{b}{\citet{caswell952} }
\tablenotetext{c}{\citet{hoefner96} }
\tablenotetext{d}{\citet{wood89} }
\end{deluxetable}

\subsection{Region G45.07+0.13}

\label{g45.07}

\begin{figure}[h!]
\centering \includegraphics[width=0.9\textwidth]{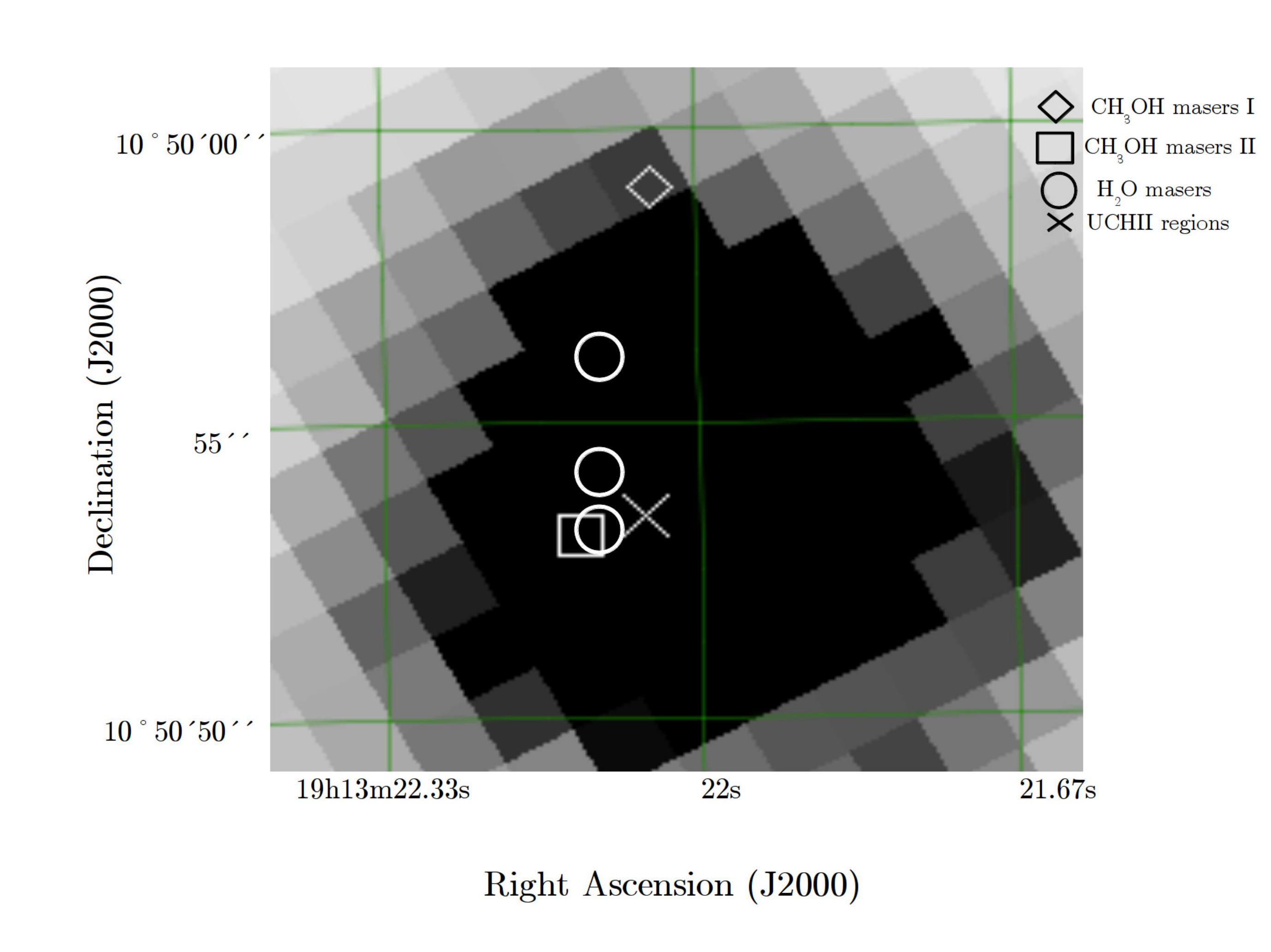}
\caption[Methanol masers and associated sources of interest in region G45.07+0.13]{Figure showing the Class II and the Class I methanol maser and other
sources of interest in region G45.07+0.13. The grayscale shows the
4.5 $\mu$m Spitzer infrared image and the intensity range is between
0.0683 MJy sr$^{-1}$ and 543 MJy sr$^{-1}$. The symbols are the
same as in Figure \ref{f9.62}}

 \label{f45.07} 
\end{figure}

Figure \ref{f45.07} shows the Class II methanol maser and the Class
I methanol maser in G45.07+0.13, together with other sources of interest.
To the south of the Class I maser position is a Class II methanol
maser, whose position is marked by a square in Figure \ref{f45.07}.
The Class I and Class II masers are separated by about $5.96^{\prime\prime}$;
at a distance of 6.0 kpc to G45.07+0.13 (\citealt{pestalozzi2005}),
this is equivalent to about 0.17 pc. A UC HII region is almost coincident
with the Class II methanol maser position, and there is 4.5~$\mu$m
emission in the entire region. The positions of these sources, together
with the telescopes used to observe them and the angular resolution
(in terms of synthesized beam) of the observations, are listed in
Table \ref{t45.07}.

\begin{figure}[h!]
\centering \includegraphics[width=0.9\textwidth]{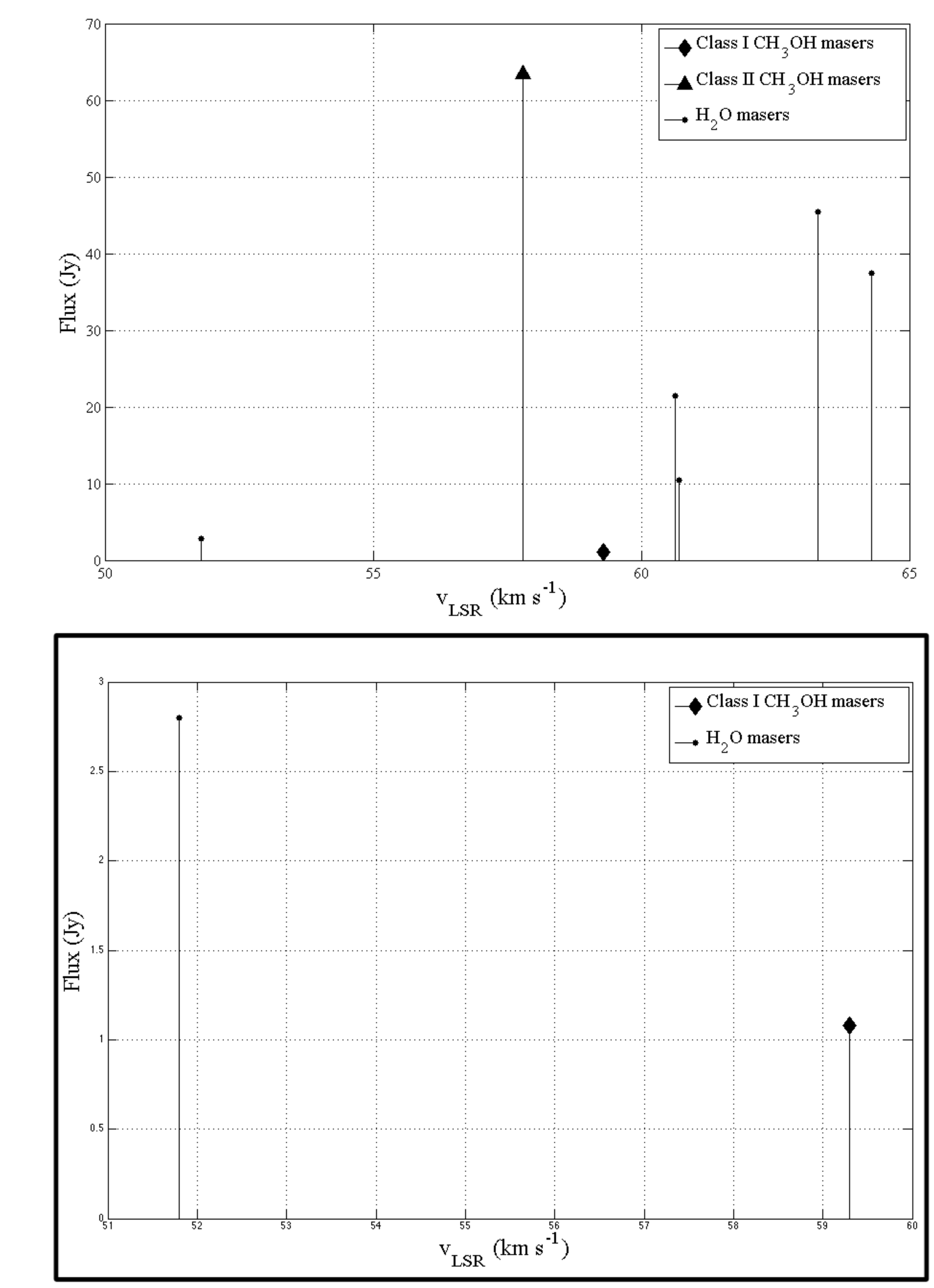}
\caption[Line profiles of methanol and H$_{2}$O masers in region G45.07+0.13]{Plot of intensities and center velocities ($v_{\text{LSR}}$) of
methanol and H$_{2}$O masers associated with G45.07+0.13. The symbols
are the same as in Figure \ref{line9.62}. The lower plot is an enlarged
version of the 0-3 Jy segment in the upper plot.}

\label{line45.07} 
\end{figure}

The center velocities ($v_{\text{LSR}}$) and intensities of the Class
I and Class II methanol masers and H$_{2}$O masers have been plotted
in Figure \ref{line45.07}. Four of the five H$_{2}$O masers shown
in this figure are at larger $v_{\text{LSR}}$ (i.e., redshifted)
than the Class II methanol maser. The Class I methanol maser is also
redshifted with respect to the Class II methanol maser.

\providecommand{\tabularnewline}{\\}
\begin{deluxetable}{ccccccc}
\tablecolumns{7}
\tablewidth{0pc}
\tabletypesize{\small}
\rotate
\tablecaption{Positions and associated information for sources associated with G45.07+0.13\label{t45.07} }
\tablehead{
\colhead{Source Type}    &\colhead{R.A. (J2000)} &\colhead{Dec. (J2000)}&\colhead{Observing}&\colhead{Synthesized} & \multicolumn{2}{c}{Kinematic}\tabularnewline
\colhead{} & \colhead{(h m s)}   & \colhead{(${^{o}}~^{\prime}~^{\prime\prime}$)}    & \colhead{Telescope} &\colhead{Beam}    & \colhead{$v_{\text{LSR}}$ (km s$^{-1}$)} &\colhead{ Intensity (Jy)}}
\startdata

44 GHz Class I CH$_{3}$OH maser \tablenotemark{a} & 19 13 22.1  & +10 50 59  & VLA  & $1.98^{\prime\prime}\times1.6^{\prime\prime}$  & 59.3 & 1.1\tabularnewline
6.7 GHz Class II CH$_{3}$OH maser \tablenotemark{b}  & 19 13 22.1  & +10 50 53 & MERLIN\tablenotemark{e} & $0.4^{\prime\prime}\times0.4^{\prime\prime}$  & 57.8 & 64\tabularnewline
22 GHz H$_{2}$O maser \tablenotemark{c} & 19 13 22.1  & +10 50 53  & VLA  & $0.61^{\prime\prime}\times0.37^{\prime\prime}$  & 60.7 & 11\tabularnewline
22 GHz H$_{2}$O maser  & 19 13 22.1  & +10 50 53  & VLA  & $0.61^{\prime\prime}\times0.37^{\prime\prime}$  & 51.8 & 2.8\tabularnewline
22 GHz H$_{2}$O maser  & 19 13 22.1  & +10 50 54  & VLA  & $2.6{}^{\prime\prime}\times0.37^{\prime\prime}$  & 63.3 & 45\tabularnewline
22 GHz H$_{2}$O maser  & 19 13 22.1  & +10 50 56  & VLA  & $0.4^{\prime\prime}\times0.37^{\prime\prime}$  & 60.6 & 22\tabularnewline
22 GHz H$_{2}$O maser  & 19 13 22.1  & +10 50 56  & VLA  & $0.4^{\prime\prime}\times0.37^{\prime\prime}$  & 64.3 & 38\tabularnewline
14.9 GHz UCHII region \tablenotemark{d} & 19 13 23.7  & +10 50 52  & VLA  & $0.80^{\prime\prime}\times0.72^{\prime\prime}$  & \nodata & \nodata\tabularnewline
 &  &  &  &  &  & \tabularnewline
\enddata
\par
\tablenotetext{a}{\citet{kurtz2004} } 
\tablenotetext{b}{\citet{caswell952} }
\tablenotetext{c}{\citet{hoefner96} }
\tablenotetext{d}{\citet{wood89} }
\tablenotetext{e}{Multi-Element Radio Linked Interferometer Network (MERLIN)}
\end{deluxetable}

\subsection{Region G45.47+0.07}

\label{g45.47}

\begin{figure}[h!]
\centering \includegraphics[width=0.9\textwidth]{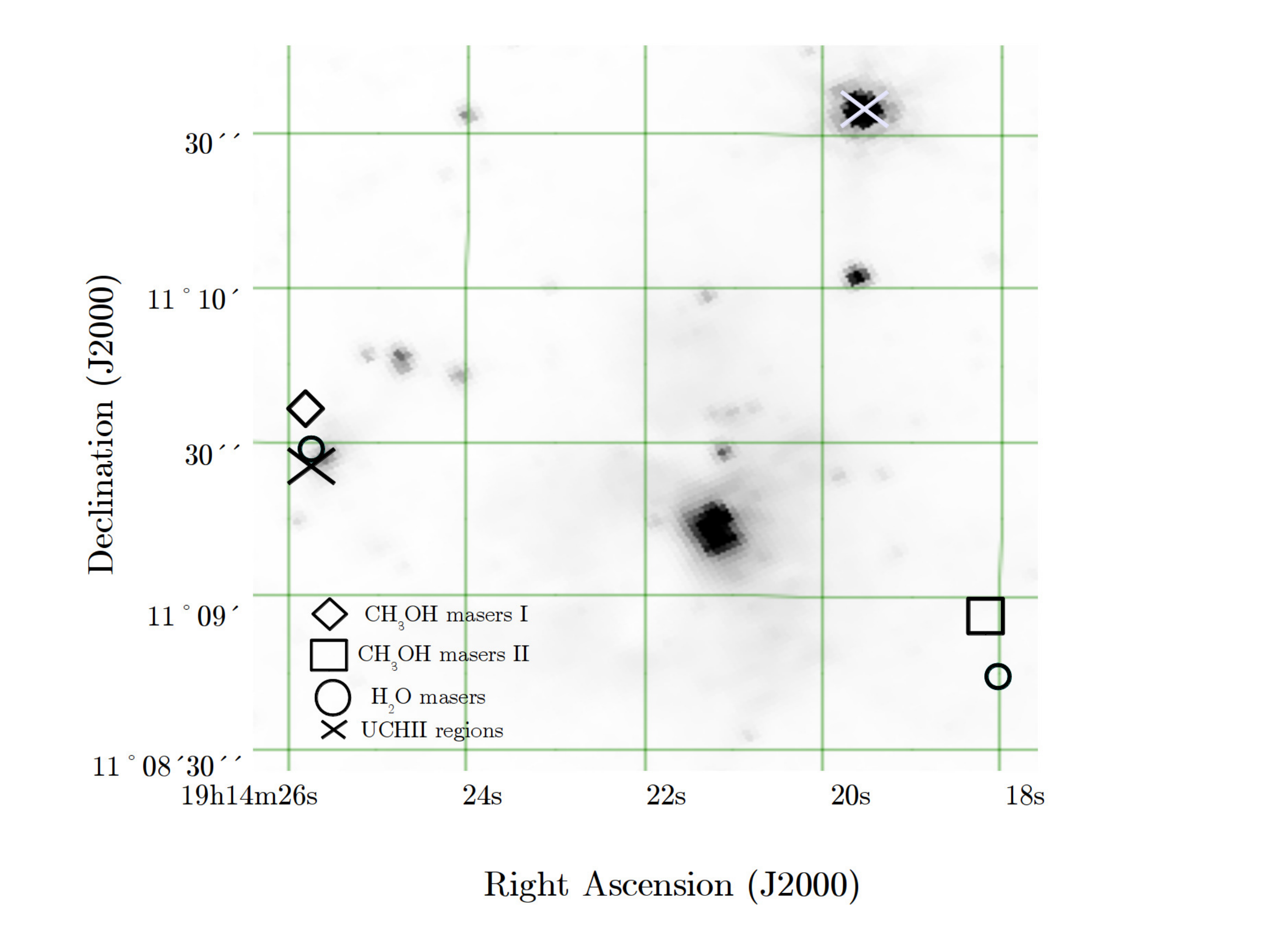}
\caption[Methanol masers and associated sources of interest in region G45.47+0.07]{Figure showing the Class II and the Class I methanol maser and other
sources of interest in region G45.47+0.07. The grayscale shows the
4.5 $\mu$m Spitzer infrared image and the intensity range is between
0.385 MJy sr$^{-1}$ and 538 MJy sr$^{-1}$. The symbols are the same
as in Figure \ref{f9.62}.}

 \label{f45.472} 
\end{figure}

Figure \ref{f45.472} shows the Class II methanol maser and the Class
I methanol maser in G45.47+0.07, together with other sources of interest.
The Class I and Class II masers are separated by about $115.2^{\prime\prime}$;
at a distance of 4.0 kpc to G45.47+0.07 (\citealt{pestalozzi2005}),
this is equivalent to about 2.23 pc. A UC HII region is $12^{\prime\prime}$
due south of the Class I methanol maser position. The 4.5~$\mu$m
infrared emission is present throughout the region, with discrete
enhancements at several locations, including one that is almost coincident
with the position of the UC HII region. The positions of these sources,
together with the telescopes used to observe them and the angular
resolution (in terms of synthesized beam) of the observations, are
listed in Table \ref{t45.47}.

\begin{figure}[h!]
\centering \includegraphics[width=0.9\textwidth]{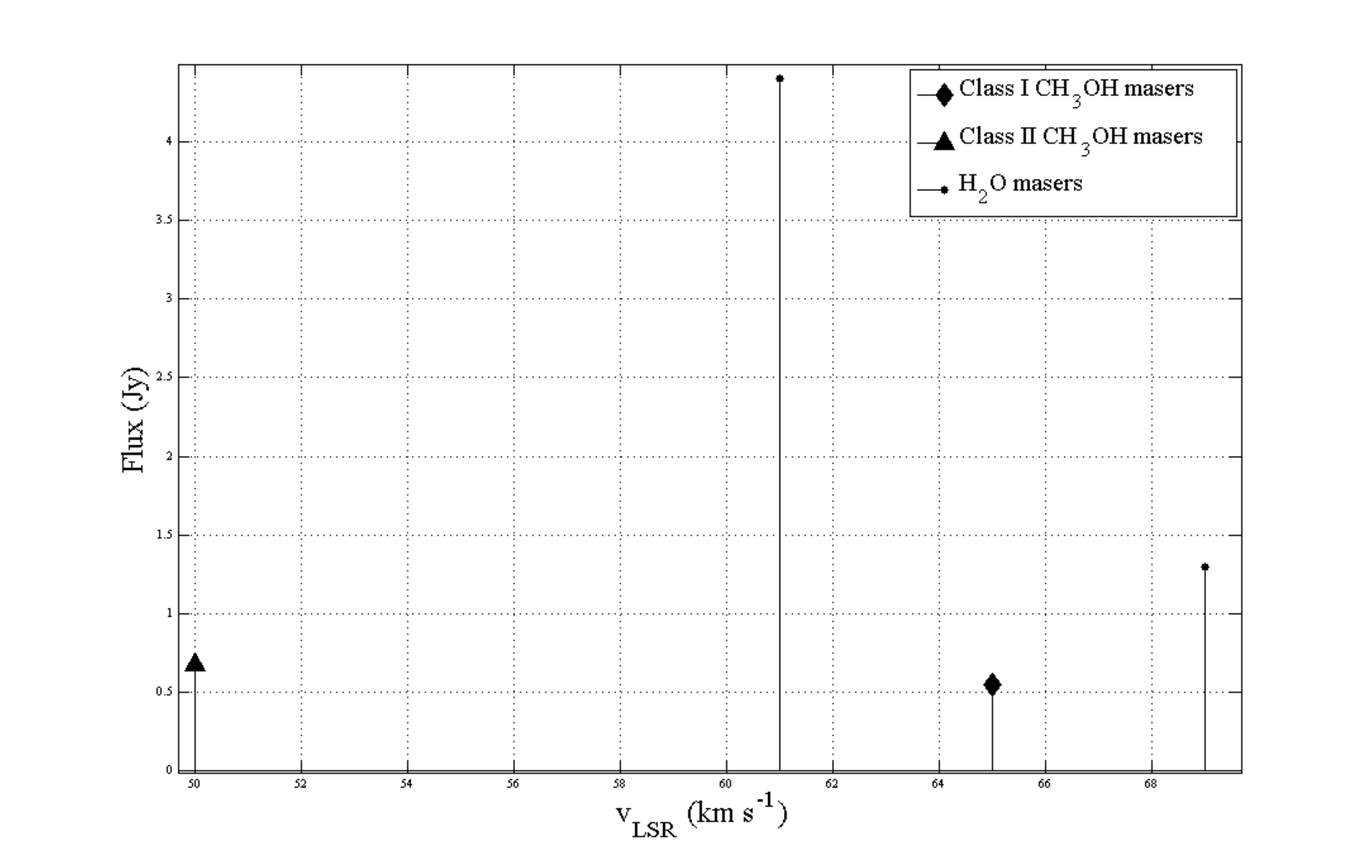}
\caption[Line profiles of methanol and H$_{2}$O masers in region G45.47+0.07]{Plot of intensities and center velocities ($v_{\text{LSR}}$) of
methanol and H$_{2}$O masers associated with G45.47+0.07. The symbols
are the same as in Figure \ref{line9.62}.}

\label{line45.47} 
\end{figure}

The center velocities ($v_{\text{LSR}}$) and intensities of the Class
I and Class II methanol masers and H$_{2}$O masers have been plotted
in Figure \ref{line45.47}. Both of the H$_{2}$O masers are at larger
$v_{\text{LSR}}$ (i.e., redshifted) than the Class II methanol maser.
The Class I methanol maser is also redshifted with respect to the
the Class II methanol maser.

\providecommand{\tabularnewline}{\\}
\begin{deluxetable}{ccccccc}
\tablecolumns{7}
\tablewidth{0pc}
\tabletypesize{\small}
\rotate
\tablecaption{Positions and associated information for sources associated with G45.47+0.07\label{t45.47}}
\tablehead{
\colhead{Source Type}    &\colhead{R.A. (J2000)} &\colhead{Dec. (J2000)}&\colhead{Observing}&\colhead{Synthesized} & \multicolumn{2}{c}{Kinematic}\tabularnewline
\colhead{} & \colhead{(h m s)}   & \colhead{(${^{o}}~^{\prime}~^{\prime\prime}$)}    & \colhead{Telescope} &\colhead{Beam}    & \colhead{$v_{\text{LSR}}$ (km s$^{-1}$)} &\colhead{ Intensity (Jy)}}
\startdata

44 GHz Class I CH$_{3}$OH maser \tablenotemark{a} & 19 14 25.7  & +11 09 37  & VLA  & $2.17^{\prime\prime}\times1.54^{\prime\prime}$  & 65 & 0.55\tabularnewline
6.7 GHz Class II CH$_{3}$OH maser \tablenotemark{b}  & 19 14 18.3  & +11 08 59  & MERLIN  & $0.1^{\prime\prime}\times0.1^{\prime\prime}$  & 50 & 0.68\tabularnewline
22 GHz H$_{2}$O maser \tablenotemark{c} & 19 14 18.1  & +11 08 47  & VLA  & $2.6^{\prime\prime}\times1.8^{\prime\prime}$  & 61  & 4.4\tabularnewline
22 GHz H$_{2}$O maser  & 19 14 25.7  & +11 09 30  & VLA  & $3.5^{\prime\prime}\times1.1^{\prime\prime}$  & 69  & 1.3\tabularnewline
5 GHz UCHII region \tablenotemark{d} & 19 14 25.7  & +11 09 26  & VLA  & $0.42^{\prime\prime}\times0.42^{\prime\prime}$  & \nodata  & \nodata\tabularnewline
14.9 GHz UCHII region \tablenotemark{e} & 19 14 19.6  & +11 10 35  & VLA  & $0.4^{\prime\prime}\times0.4^{\prime\prime}$  & \nodata  & \nodata\tabularnewline
 &  &  &  &  &  & \tabularnewline
\enddata
\par
\tablenotetext{a}{\citet{kurtz2004} } 
\tablenotetext{b}{\citet{pandian2011} }
\tablenotetext{c}{\citet{hoefner96} }
\tablenotetext{d}{\citet{urquhart2009}}
\tablenotetext{e}{\citet{wood89} }
\end{deluxetable}

\subsection{Region G75.78+0.34}

\label{g75.78}

\begin{figure}[h!]
\centering \includegraphics[width=0.9\textwidth]{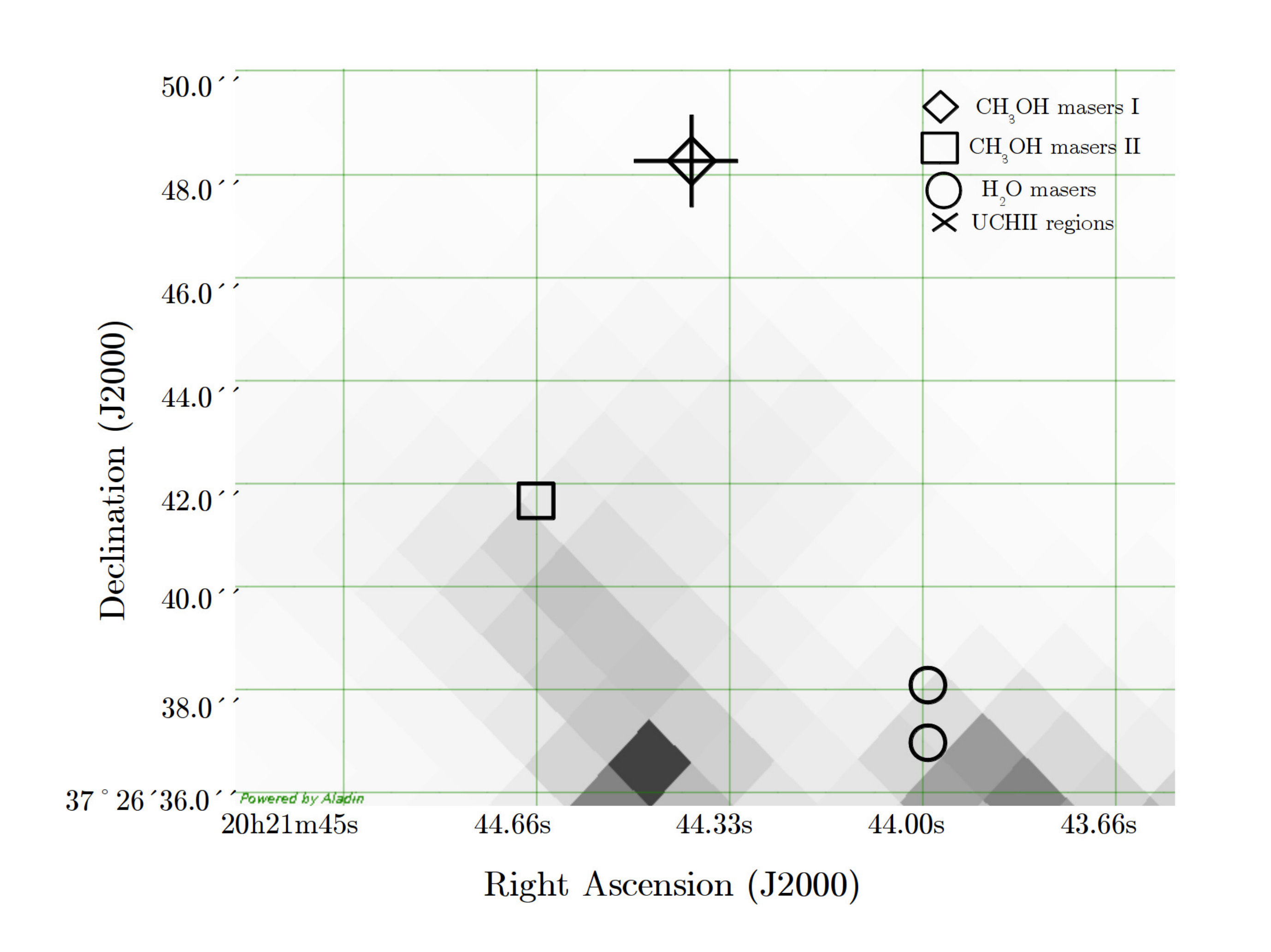}
\caption[Methanol masers and associated sources of interest in region G75.78+0.34]{Figure showing the Class II and the Class I methanol maser and other
sources of interest in region G75.78+0.34. The grayscale shows the
4.5 $\mu$m Spitzer infrared image and the intensity range is between
0.536 MJy sr$^{-1}$ and 797 MJy sr$^{-1}$. The symbols are the same
as in Figure \ref{f9.62}.}

 \label{f75.78} 
\end{figure}

Figure \ref{f75.78} shows the Class II methanol maser and the Class
I methanol maser in G75.78+0.34, together with other sources of interest.
The Class I and Class II masers are separated by about $7.12^{\prime\prime}$;
at a distance of 4.9 kpc to G75.78+0.34 (\citealt{pestalozzi2005}),
this is equivalent to about 0.17 pc. Two H$_{2}$O masers lie $10.0^{\prime\prime}$
to the southwest of the Class I methanol maser. The positions of these
sources, together with the telescopes used to observe them and the
angular resolution (in terms of synthesized beam) of the observations,
are listed in Table \ref{t75.78}.

\begin{figure}[h!]
\centering \includegraphics[width=0.9\textwidth]{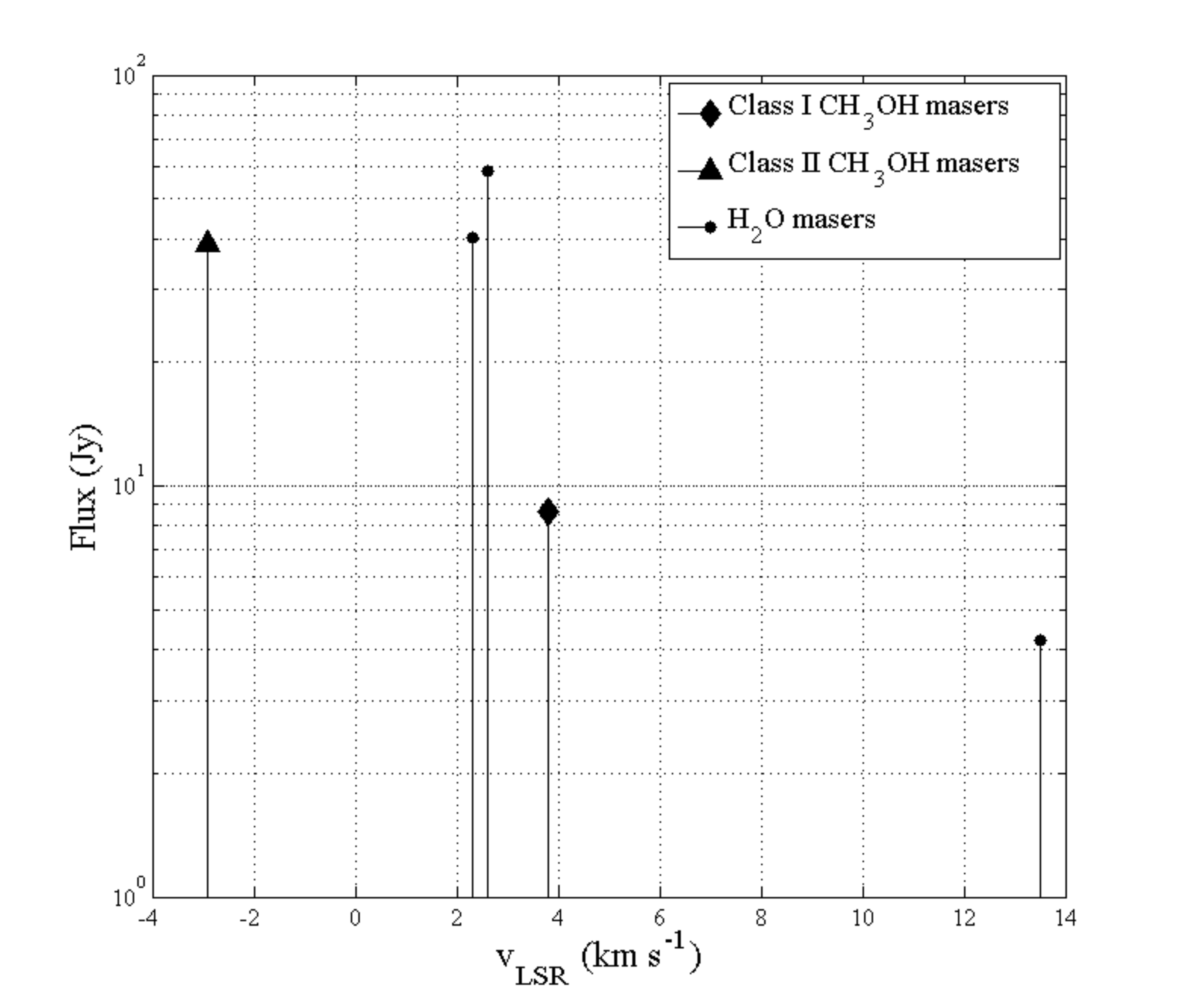}
\caption[Line profiles of methanol and H$_{2}$O masers in region G75.78+0.34]{Plot of intensities and center velocities ($v_{\text{LSR}}$) of
methanol and H$_{2}$O masers associated with G75.78+0.34. The symbols
are the same as in Figure \ref{line9.62}.}

\label{line75.78} 
\end{figure}

The center velocities ($v_{\text{LSR}}$) and intensities of the Class
I and Class II methanol masers and H$_{2}$O masers have been plotted
in Figure \ref{line75.78}. Two of the three H$_{2}$O masers are
at smaller $v_{\text{LSR}}$ (i.e., blueshifted) than the Class II
methanol maser. The Class I methanol maser is also blueshifted with
respect to the Class II methanol maser.

\providecommand{\tabularnewline}{\\}
\begin{deluxetable}{ccccccc}
\tablecolumns{7}
\tablewidth{0pc}
\tabletypesize{\small}
\rotate
\tablecaption{Positions and associated information for sources associated with G75.78+0.34\label{t75.78} }
\tablehead{
\colhead{Source Type}    &\colhead{R.A. (J2000)} &\colhead{Dec. (J2000)}&\colhead{Observing}&\colhead{Synthesized} & \multicolumn{2}{c}{Kinematic}\tabularnewline
\colhead{} & \colhead{(h m s)}   & \colhead{(${^{o}}~^{\prime}~^{\prime\prime}$)}    & \colhead{Telescope} &\colhead{Beam}    & \colhead{$v_{\text{LSR}}$ (km s$^{-1}$)} &\colhead{ Intensity (Jy)}}
\startdata

44 GHz Class I CH$_{3}$OH maser \tablenotemark{a}  & 20 21 44.4  & +37 26 48 & VLA  & $1.5^{\prime\prime}\times1.5^{\prime\prime}$  & 3.8 & 8.7\tabularnewline
6.7 GHz Class II CH$_{3}$OH maser \tablenotemark{b}   & 20 21 44.7  & +37 26 42  & VLA  & $0.5^{\prime\prime}\times0.5^{\prime\prime}$  & -2.9  & 39\tabularnewline
22 GHz H$_{2}$O maser \tablenotemark{c}  & 20 21 44.0  & +37 26 38  & VLA  & $0.52^{\prime\prime}\times0.3^{\prime\prime}$  & 2.6  & 58\tabularnewline
22 GHz H$_{2}$O maser  & 20 21 44.0  & +37 26 37  & VLA  & $0.52^{\prime\prime}\times0.3^{\prime\prime}$  & 14  & 4.2\tabularnewline
22 GHz H$_{2}$O maser  & 20 21 44.0  & +37 26 37  & VLA  & $0.52^{\prime\prime}\times0.3^{\prime\prime}$  & 2.3  & 40\tabularnewline
 &  &  &  &  &  & \tabularnewline
\enddata
\par
\tablenotetext{a}{\citet{kurtz2004} } 
\tablenotetext{b}{\citet{caswell952} }
\tablenotetext{c}{\citet{hoefner96} }
\end{deluxetable}

\subsection{Regions for which high resolution data were only available for Class
I and Class II methanol masers.}

\label{lowres}

In Figure \ref{piechart}, we saw that for 60\% of the regions we
examined, high resolution data were present for the Class I and Class
II methanol masers only, but not for the other associated tracers
of star formation discussed in \S\ \ref{discussion}. The data on
these methanol masers for such regions are presented in Table \ref{tlowres},
including coordinates of each region, the offset in pc between the
Class I and Class II methanol masers, and the characteristics of the
spectral lines of each maser.

The projected distance between Class I and Class II methanol masers
in these twenty-two regions is in general agreement with the separation
between masers discussed in \S\ \ref{g9.62} to \S\ \ref{g75.78}.
The maximum angular separation is 59.34$^{\prime\prime}$ for G24.943+0.074;
at a distance of 7.7 kpc (\citealt{pestalozzi2005}), this works out
to 2.2 pc, similar to the projected distance between Class I and Class
II masers in G45.47+0.07. The minimum projected separation was zero
in three cases: G28.28-0.36, G37.47-0.11, G49.27-0.34, but that is
likely due to the resolution of the observations not being great enough
to resolve the Class I and Class II methanol maser sources. The mean
projected separation between Class I and Class II methanol masers
for all thirty regions discussed in this paper was 0.67 pc with a
root mean squared error of 0.12 pc.

\providecommand{\tabularnewline}{\\}
\begin{deluxetable}{cccccccc}
\tablecolumns{8}
\tablewidth{0pc}
\tabletypesize{\small}
\rotate
\tablecaption{Regions with high resolution data for methanol masers only\label{tlowres}}
\tablehead{
\colhead{Region}  & \colhead{R.A. (J2000)}  & \colhead{Dec. (J2000)} & \colhead{MMI MMII offset \tablenotemark{e}}  &\multicolumn{2}{c}{ 44 GHz Class I $CH_{3}OH$}  & \multicolumn{2}{c}{6.7 GHz Class II $CH_{3}OH$ \tablenotemark{c}}  \tabularnewline
\colhead{}& \colhead{(h m s)}  & \colhead{(${^{o}}~^{\prime}~^{\prime\prime}$)} & \colhead{(pc)} & \colhead{$v_{\text{LSR}}$(km s$^{-1}$)}   & \colhead{Intensity (Jy)}    &  \colhead{$v_{\text{LSR}}$(km s$^{-1}$) } & \colhead{ Intensity (Jy)}\tabularnewline}

\startdata
G13.66-0.60 \tablenotemark{a} & 18 17 22.9  & -17 22 13.8  & 0.84$\pm$0.02  & 48  & 28  & 51  & 33\tabularnewline
G19.36-0.02 \tablenotemark{b} & 18 26 25.8  & -12 03 56.9  & 0.96$\pm$0.06  & 27  & 54  & 26  & 6.7\tabularnewline
G43.04-0.45 \tablenotemark{a} & 19 11 38.8  & 08 46 37.9  & 0.67$\pm$0.01  & 58  & 3.8  & 55  & 10\tabularnewline
G94.26-0.41 \tablenotemark{a} & 21 32 30.7  & 51 02 14.9  & 0.31$\pm$0.01  & -48  & 1.1  & -48  & 1.1\tabularnewline
G10.29-0.13 \tablenotemark{b} & 18 08 49.5  & -20 05 53.5  & 0.35$\pm$0.01  & 14  & 2.0  & 10  & 2.7\tabularnewline
G14.89-0.40 \tablenotemark{a} & 18 19 7.60  & -16 11 25.6  & 0.61$\pm$0.01  & 62  & 7.8  & 63  & 3.2\tabularnewline
G18.66+0.04 \tablenotemark{b} & 18 24 53.7  & -12 39 20.0  & 1.6$\pm$0.02  & 80  & 2.1  & 80  & 22\tabularnewline
G18.89-0.47 \tablenotemark{b} & 18 27 7.90  & -12 41 35.5  & 0.01$\pm$0.01  & 66  & 20  & 38  & 3.7\tabularnewline
G22.04+0.22 \tablenotemark{b} & 18 30 34.7  & -09 34 47.0  & 0.97$\pm$0.02  & 51  & 32  & 54  & 6.1\tabularnewline
G23.96-0.11 \tablenotemark{a} & 18 35 22.3  & -08 01 28.0  & 0.72$\pm$0.02  & 73  & 6.2  & 71  & 17\tabularnewline
G24.94+0.07 \tablenotemark{b} & 18 36 31.5  & -07 04 16.0  & 2.2$\pm$0.02  & 41  & 1.7  & 53  & 4.2\tabularnewline
G25.27-0.43 \tablenotemark{b} & 18 38 56.9  & -07 00 48.0  & 0.03$\pm$0.01  & 60  & 6.2  & 59  & 0.55\tabularnewline
G28.28-0.36 \tablenotemark{b} & 18 44 13.2  & -04 18 4.00  & 0.00$\pm$0.02  & 48  & 4.1  & 81  & 62\tabularnewline
G28.84-0.25 \tablenotemark{b} & 18 44 51.0  & -03 45 49.0  & 1.1$\pm$0.02  & 87  & 5.6  & 100  & 1.9\tabularnewline
G34.82+0.35 \tablenotemark{a} & 18 53 37.7  & 01 50 25.4  & 0.31$\pm$0.01  & 57  & 0.67  & 56  & 0.27\tabularnewline
G36.11+0.55 \tablenotemark{a} & 18 55 16.8  & 03 05 6.70  & 1.0$\pm$0.01  & 76  & 0.57  & 73  & 19\tabularnewline
G37.27+0.08 \tablenotemark{a} & 18 59 3.73  & 03 53 42.9  & 0.88$\pm$0.01  & 90  & 1.0  & 90  & \nodata\tablenotemark{d}\tabularnewline
G37.47-0.11 \tablenotemark{b} & 19 00 7.00  & 03 59 53.0  & 0.00$\pm$0.01  & 59  & 0.31  & 58  & 13\tabularnewline
G39.10+0.48\tablenotemark{b} & 19 00 58.1  & 05 42 44.0  & 0.62$\pm$0.01  & 27  & 11  & 15  & 15\tabularnewline
G49.27-0.34 \tablenotemark{b} & 19 23 6.70  & 14 20 13.0  & 0.00$\pm$0.01  & 67  & 0.65  & 67  & \nodata\tablenotemark{d}\tabularnewline
 &  &  &  &  &  & &\tabularnewline
\enddata
\par

\tablenotetext{a}{\citet{kurtz2004}}

\tablenotetext{b}{\citet{cyganowski2009}}

\tablenotetext{c}{\citet{pestalozzi2005}}

\tablenotetext{d}{\citet{pestalozzi2005} reports a position fix only.}

\tablenotetext{e}{The values in this column were computed using a measurement of
the distance from our sun to the region in question, the error indicates
the propagated uncertainty in the determination of the heliocentric
distance.}

\tablecomments{For a valid comparison across sources, the data has been rounded
to the least number of significant figures.}
\end{deluxetable}

\clearpage{}

\section{Discussion}

\label{Chapter4}

In this section, we discuss the morphology and kinematics of the eight
regions for which high resolution data were available. In \S\ \ref{morphkin},
we identify three patterns of morphology and a notable trend in the
center velocities of the masers. In \S\ \ref{models} we describe
three physical models which can explain the patterns discussed in
\S\ \ref{morphkin}.

\subsection{Morphology and Kinematics\label{morphkin}}

Of the thirty regions that we found to have data on Class I and Class
II methanol masers, we were able to obtain high resolution data for
associated star formation tracers for eight regions. Such associated
data include H$_{2}$O masers, ultracompact (UC) ionized hydrogen
(H II) regions, and 4.5 $\mu$m infrared data from Spitzer. For these
eight regions, we looked at the morphology of Class I and II methanol
masers, and H$_{2}$O masers, and the center velocities of the line
profiles of these masers. We also looked at the location of the UC
H II regions and 4.5 $\mu$m infrared sources. Based on our examination
of these data, we have identified three distinct patterns in the morphology
of the masers, and one notable trend in the center velocities of the
Class I methanol and H$_{2}$O maser spectral lines in relation to
the center velocity of the Class II methanol masers.

\subsection{Linear arrangement of H$_{2}$O masers\label{H2linear}}

In two of the eight sources, G9.62+0.19 and G45.07+0.13, the H$_{2}$O
masers are in a linear arrangement, spread out along a straight line
(approximately) between the Class I and Class II methanol maser (Figure~\ref{f9.62}
and \ref{f45.07}). There are nine H$_{2}$O masers in G9.62+0.19,
but only five in G45.07+0.13, so the linear structure is more pronounced
in G9.62+0.19. The projected separation between the two classes of
methanol maser is 0.27 pc for G9.62+0.19 and 0.17 pc for G45.07+0.13.
A UC H II region is almost coincident with the Class II methanol maser
in both sources. Since UC H II regions are formed by hot and massive
young protostars, it is reasonable to conclude that the same ionizing
radiation that is producing the UC H II region is also likely providing
the radiative excitation of the Class II methanol maser. The linear
arrangement of H$_{2}$O masers between the Class II and Class I methanol
masers in both G9.62+0.19 and G45.07+0.13 is also suggestive of an
outflow, with the Class I methanol maser excited near the head of
the outflow in both sources. As stated in \S\ \ref{g9.62}, we consider
masers with $v_{\text{LSR}}$ larger than the $v_{\text{LSR}}$ of
the Class II methanol maser to be redshifted. Seven of the nine H$_{2}$O
masers in G9.62+0.19 are redshifted with respect to the Class II methanol
maser (e.g., see Table \ref{t9.62}, where the $v_{\text{LSR}}$ of
7 H$_{2}$O masers are larger than the $v_{\text{LSR}}$ of the Class
II methanol maser, and Figure~\ref{line9.62}). Therefore, these
7 masers are likely in the redshifted lobe of the outflow (see \S\ \ref{DOM1}
for additional discussion). The Class I methanol maser also has a
larger $v_{\text{LSR}}$ than the $v_{\text{LSR}}$ of the Class II
maser, so it can also be located in the redshifted lobe of the outflow.
Likewise, in G45.07+0.13, three out of four masers are redshifted
with respect to the Class II methanol maser, as is the Class I methanol
maser (Table~\ref{t45.07} and Figure~\ref{line45.07}).

In G9.62+0.19, however, we cannot rule out that the Class I methanol
maser may not be associated with the Class II methanol maser, particularly
given the recent discovery by \citet{Voronkov2010MNRAS} that suggests
that Class I masers may also be excited in shocks driven by expanding
H II regions. Indeed, there is a UC H II region to the southwest of
the Class I methanol maser in G9.62+0.19 and a 4.5 $\mu$m infrared
source near the UC H II region, indicating another center of high
mass star formation activity at this location (Figure~\ref{f9.62}).
Still, with 7 of 9 H$_{2}$O masers in G9.62+0.19 redshifted with
respect to the Class II maser, the outflow interpretation is reasonable,
so that the data are certainly compatible with disk-outflow systems.

\subsection{H$_{2}$O masers clustered near Class II methanol maser\label{H2nearClassII}}

In two of the eight regions, G10.47+0.03 and G75.78+0.34, the H$_{2}$O
masers are far away from the Class I methanol maser (Figure~\ref{f10.47}
and \ref{f75.78}). In G10.47+0.03, all the three H$_{2}$O masers
are near the Class II methanol maser. A faint 4.5 $\mu$m source and
a UC H II region are coincident with the Class II and H$_{2}$O maser
position. As stated in \S\ \ref{g9.62}, we consider masers with
$v_{\text{LSR}}$ smaller than the $v_{\text{LSR}}$ of the Class
II methanol maser to be blueshifted. All three H$_{2}$O masers in
G10.47+0.03, and the Class I methanol maser, are blueshifted with
respect to the Class II methanol maser (Table~\ref{t10.47} and Figure~\ref{line10.47}).

In G75.78+0.34, all three H$_{2}$O masers are at $\sim$ 9.1 pc from
the Class II methanol maser, but a line from the Class II methanol
maser to the H$_{2}$O masers is roughly perpendicular to a line joining
the Class I and II methanol masers (Figure~\ref{f75.78}). An extended
4.5 $\mu$m source straddles the Class II methanol maser and H$_{2}$O
maser positions. All three H$_{2}$O masers, and the Class I methanol
maser, are redshifted with respect to the Class II methanol maser
(Table~\ref{t75.78} and Figure~\ref{line75.78}). We speculate
that the Class II methanol maser and H$_{2}$O masers are located
in a circumstellar disk in both these sources, and discuss this further
in \S\ \ref{DOM3}.

\subsection{H$_{2}$O masers clustered near Class I methanol maser\label{H2nearClassI}}

In three of the eight sources, G12.20-0.11, G31.41+0.31, and G35.03+0.35,
the H$_{2}$O masers are far away from the Class II methanol maser,
but are located near the Class I methanol maser (Figures~\ref{f12.20},
\ref{f31.41}, and \ref{f35.03}). Of the 6 masers near the Class
I methanol maser in G12.20-0.11, two H$_{2}$O masers, and the Class
I methanol maser, are redshifted with respect to the Class II methanol
maser (Table~\ref{t12.20} and Figure~\ref{line12.20}). Meanwhile,
three H$_{2}$O masers are blueshifted, and one has the same center
velocity, as the Class II methanol maser. Of the 8 masers in G31.41+0.31,
five H$_{2}$O masers, and the Class I methanol maser, are blueshifted
with respect to the Class II methanol maser, whereas two H$_{2}$O
masers are redshifted, and one has the same $v_{\text{LSR}}$ as the
Class II methanol maser (Table~\ref{t31.41} and Figure~\ref{line31.41}).
There is only one H$_{2}$O maser in G35.03+0.35, and it, along with
the Class I methanol maser, is redshifted with respect to the Class
II water maser (Table~\ref{t35.03} and Figure~\ref{line35.03}).
In all the three regions, G12.20-0.11, G31.41+0.31, and G35.03+0.35,
there is a 4.5 $\mu$m source that is closer to the Class I methanol
maser than the Class II methanol maser. In \S\ \ref{DOM2}, we will
ascribe the geometry in these sources to one or more outflows.

\subsection{The eighth source G45.47+0.07 \label{source8}}

The eighth source for which we have high angular resolution data for
H$_{2}$O masers and other associated tracers, G45.47+0.07, does not
fit into any of the three categories listed above. It has one H$_{2}$O
maser near the Class I methanol maser source, and one H$_{2}$O maser
near the Class II methanol maser source (Figure~\ref{f45.472}).
The Class I methanol maser, an H$_{2}$O maser near it, and the H$_{2}$O
maser near the Class II methanol maser, are all redshifted with respect
to the Class II methanol maser (Table~\ref{t45.47} and Figures~\ref{line45.47}).
We interpret this as a hybrid of the models discussed in \S\ \ref{H2nearClassII}
and \S\ \ref{H2nearClassI}, and discuss it further in \S\ \ref{DOM4}.

\section{Models based on high angular resolution data. \label{models}}

\begin{figure}[h!]
\centering \includegraphics[width=1\textwidth]{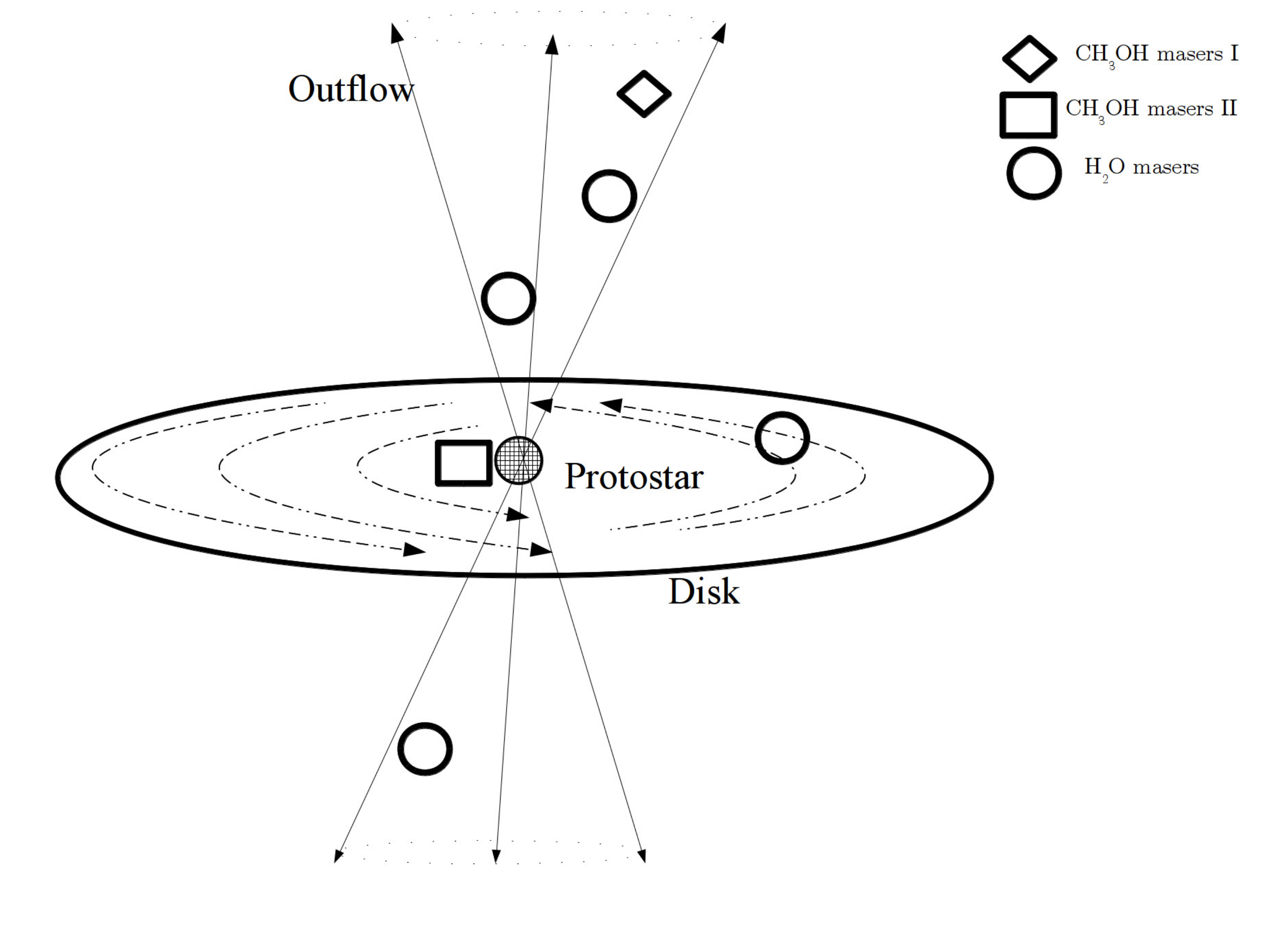}
\caption[Diagram showing an accretion disk and bipolar outflow from a high
mass protostar.]{Diagram showing an accretion disk and outflow from a massive young
protostar. As discussed in \S\ \ref{Chapter1}, star formation
is accompanied by a circumstellar disk from which the central protostar
accretes material, and a bipolar outflow at right angles to the disk
which carries mass and angular momentum away from the protostar. Using
the same symbols as were used in \S\ \ref{Chapter3} we have shown
some typical locations for Class I and Class II methanol masers and
water masers in relation to the circumstellar disk, outflow and massive
protostar. In general, the disk may be inclined at some angle to the
line of sight, in which case one lobe of the outflow would be redshifted
and the other blueshifted.}

\label{badartist} 
\end{figure}

Based on our findings (\S\ \ref{Chapter3}) and discussion in
\S\ \ref{morphkin}, we have constructed three disk-outflow models
for methanol masers in star forming regions. As discussed in \S\
\ref{Chapter1}, star formation is accompanied by a circumstellar
disk from which the central protostar accretes material, and a bipolar
outflow at right angles to the circumstellar disk which carries mass
and angular momentum away from the protostar (Figure \ref{badartist}).
In general, the disk may be inclined at some angle to the line of
sight, in which case one lobe of the outflow would be redshifted and
the other blueshifted. We will now discuss each model in detail.

\clearpage{}

\subsection{Disk-outflow model 1\label{DOM1}}

\begin{figure}[h!]
\centering \includegraphics[width=0.9\textwidth]{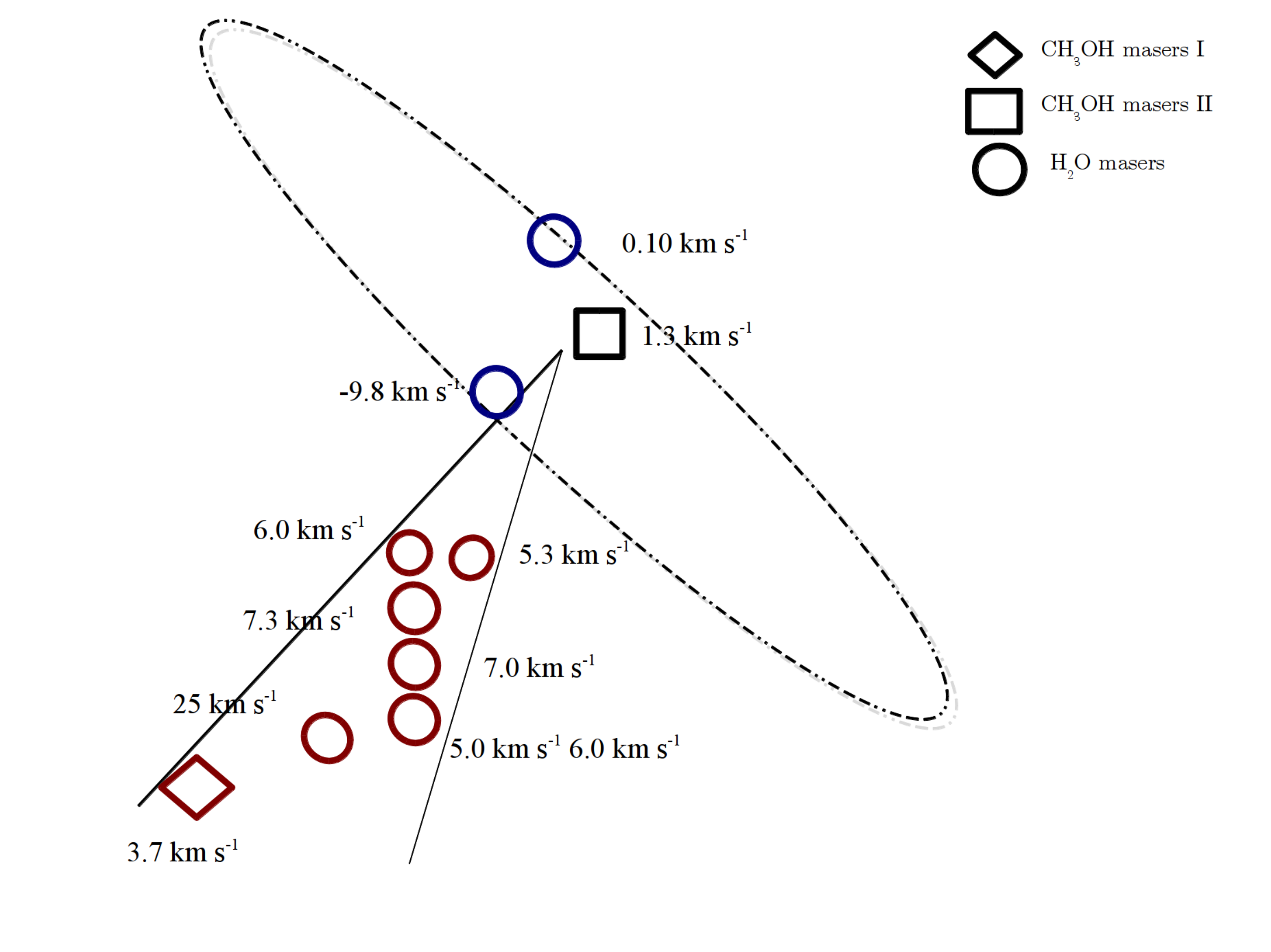}
\caption[Diagram of our disk-outflow model 1 with H$_{2}$O masers in a linear
configuration.]{Diagram of our disk-outflow model 1, showing the locations of the
Class II and I methanol masers and the H$_{2}$O masers. We have chosen
to put the information from the G9.62+0.19 star forming region (Figure~\ref{f9.62}
and Table \ref{t9.62}) in this figure. The symbols are the same as
those used in \S\ \ref{Chapter3}, and are marked at the top right
of the figure. Masers that are redshifted with respect to the Class
II methanol maser are shown in red.}

\label{linearH$_2$Oflow} 
\end{figure}

Our first model is based on the morphology of methanol and H$_{2}$O
masers in G9.62+0.19 and G45.07+0.13 (\S\ \ref{g9.62}, \S\ \ref{g45.07},
and \S\ \ref{H2linear}). Figure \ref{linearH$_2$Oflow} shows the
model of the disk and outflow, together with the locations of the
Class II and Class I methanol masers, and the H$_{2}$O masers. We
have chosen to put the information from the G9.62+0.19 star forming
region (\S\ \ref{g9.62}) in the figure. Since it is known that
Class II methanol masers are usually found near protostars, we have
put the disk near the Class II methanol maser source. Next, it is
well known that H$_{2}$O masers are formed in both outflows and circumstellar
disks in high mass star forming regions. Except for the two water
masers nearest to the Class II methanol maser in G9.62+0.19, all the
other 7 masers are redshifted (i.e., at higher center velocities,
$v_{\text{LSR}}$) with respect to the Class II methanol maser, and
well separated from it in position, as is the Class I methanol maser.
Therefore, we interpret all of them to be part of the redshifted lobe
of the outflow. We have used the position of the H$_{2}$O maser on
the extreme right and the Class I methanol maser on the left in G9.62+0.19
to constrain the edges of the redshifted outflow lobe, giving us a
nominal opening angle of 23$^{\circ}$ for the outflow. Choosing the
edges of the two redshifted H$_{2}$O masers in G45.07+0.13 also gives
us a similar opening angle for the outflow. In both regions, therefore,
the outflow appears to be well collimated. This is usually the case
in low mass star forming regions in the earlier stages of star formation,
and has been known to be true in high mass star forming regions whenever
outflows can be disentangled between sources (\citealt{McKee2007}).
In other words, due to the multiplicity of sources in high mass star
forming regions, it is usually difficult to tell which outflow is
coming from which source. Here, however, if all the redshifted masers
are truly part of the same redshifted outflow lobe that is centered
near the Class II methanol maser location, then the smaller value
of the collimation angle that we have obtained is as expected. For
example, in low mass star forming regions, \citet{Arce2007} described
collimation angles of a few degrees, and in high mass star forming
regions, \citet{Beuther2005} found outflows collimated to the same
degree as in low mass star forming regions. The two blueshifted H$_{2}$O
masers in Figure~\ref{linearH$_2$Oflow} near the Class II methanol
maser in G9.62+0.19 may be part of the blueshifted lobe of the outflow
or may be part of the circumstellar disk of the protostar. We favor
the latter interpretation for the southeastern blue shifted maser,
at least, since the blueshifted lobe should be opposite the redshifted
lobe. Likewise the redshifted and blueshifted masers that are almost
coincident with the Class II methanol maser in G45.07+0.13 (Figure~\ref{f45.07}
and Table~\ref{t45.07}) could be interpreted as being at the base
of the redshifted and blueshifted outflow lobes, or they could be
in the circumstellar disk around the protostar.

While the model of the disk and the redshifted lobe given above is
one interpretation of the data, it is worth noting that the morphology
of methanol and H$_{2}$O masers in G9.62+0.19 in particular, does
support other geometries that we have not considered here. For example,
\citet{Voronkov2010MNRAS} have suggested that some Class I methanol
masers may be excited in shocks driven by expanding H II regions.
We note that there is a UC H II region to the southwest of the Class
I methanol maser in G9.62+0.19 (Figure~\ref{f9.62}) which could
be responsible for the shock causing the Class I methanol maser.

\subsection{Disk-outflow model 2 \label{DOM3}}

\begin{figure}[h!]
\centering \includegraphics[width=0.9\textwidth]{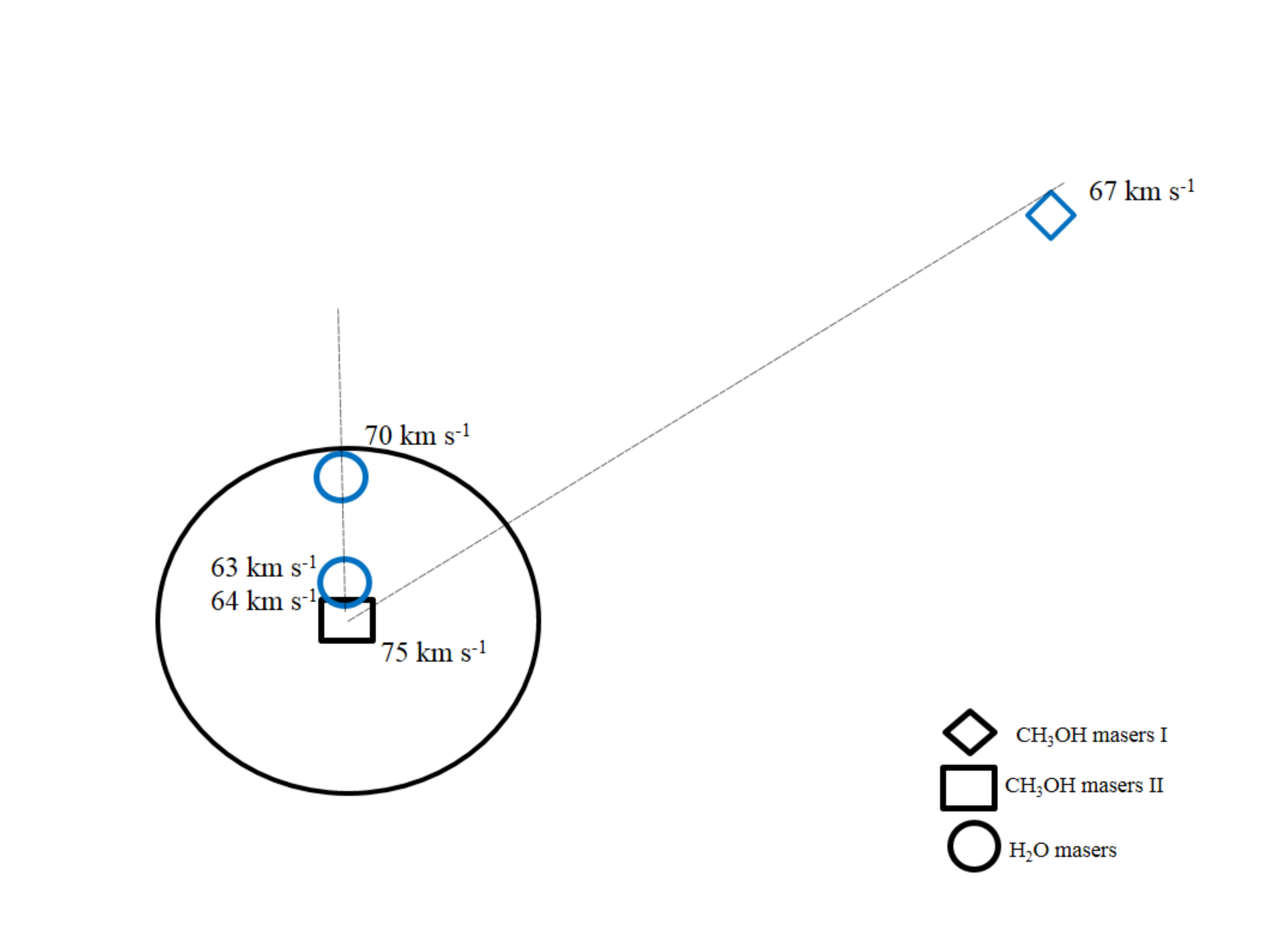}
\caption[Diagram of our disk-outflow model 2 with the Class II methanol maser
and water masers in a circumstellar disk, and the Class I maser in
the outflow.]{Diagram of our disk-outflow model 2, showing the locations of the
Class II and I methanol masers and the H$_{2}$O masers. We have chosen
to put the information from the G10.47+0.03 star forming region (Figure~\ref{f10.47}
and Table \ref{t10.47}) in this figure. The symbols are the same
as those used in \S\ \ref{Chapter3}, and are marked at the bottom
right of the figure. Masers that are blueshifted with respect to the
Class II methanol maser are shown in blue.}

\label{diskfig} 
\end{figure}

The second model is based on the morphology of methanol and H$_{2}$O
masers in G10.47+0.03 and G75.78+0.34 (\S\ \ref{g10.47}, \S\ \ref{g75.78},
and \S\ \ref{H2nearClassII}). Figure~\ref{diskfig} shows the
model of the disk and outflow, together with the locations of the
Class II and Class I methanol masers, and the H$_{2}$O masers. We
have chosen to put the information from G10.47+0.03 (\S\ \ref{g10.47})
into Figure~\ref{diskfig}. In both these sources, we interpret the
H$_{2}$O masers to be located in a circumstellar disk together with
the Class II methanol maser, with the Class I methanol maser located
in the outflow. This is a reasonable interpretation for ,
where the H$_{2}$O masers are clustered close to the Class II methanol
maser (Figure~\ref{f10.47}). In G75.78+0.34, however, while the
geometry suggests such an interpretation (Figure~\ref{f75.78}),
the size of the circumstellar disk would be too large, if it were
to be constrained by the distance between the H$_{2}$O masers and
the Class II methanol maser, which is 0.2 pc $\equiv$ 40, 000 AU
(e.g., compare to the circumstellar disk observed by \citet{{Torrelles1996}}
in Cepheus A with a 300 AU radius). Therefore, it is more likely that
either the Class I and II methanol masers are associated in G75.78+0.34,
and the H$_{2}$O masers are part of an outflow or disk of another
system (e.g., there is a 4.5 $\mu$m peak near the H$_{2}$O masers),
or that the Class II methanol and H$_{2}$O masers are associated,
and the Class I methanol maser is part of an outflow from a nearby source.
Recently \citet{Titmarsh2013} observed this region with the ATCA working 
at 22 Ghz and found strong evidence of a high velocity ( $v_{\text{LSR}}$ approaching 100 km s$^{-1}$) outflow.  \citet{Titmarsh2013} was published after the thesis (\citealt{Farmer2013}) on which this paper is based was defended,  while this paper was in preparation, and therefore their data was not used.

\subsection{Disk-outflow model 3 \label{DOM2}}

\begin{figure}[h!]
\centering \includegraphics[width=0.9\textwidth]{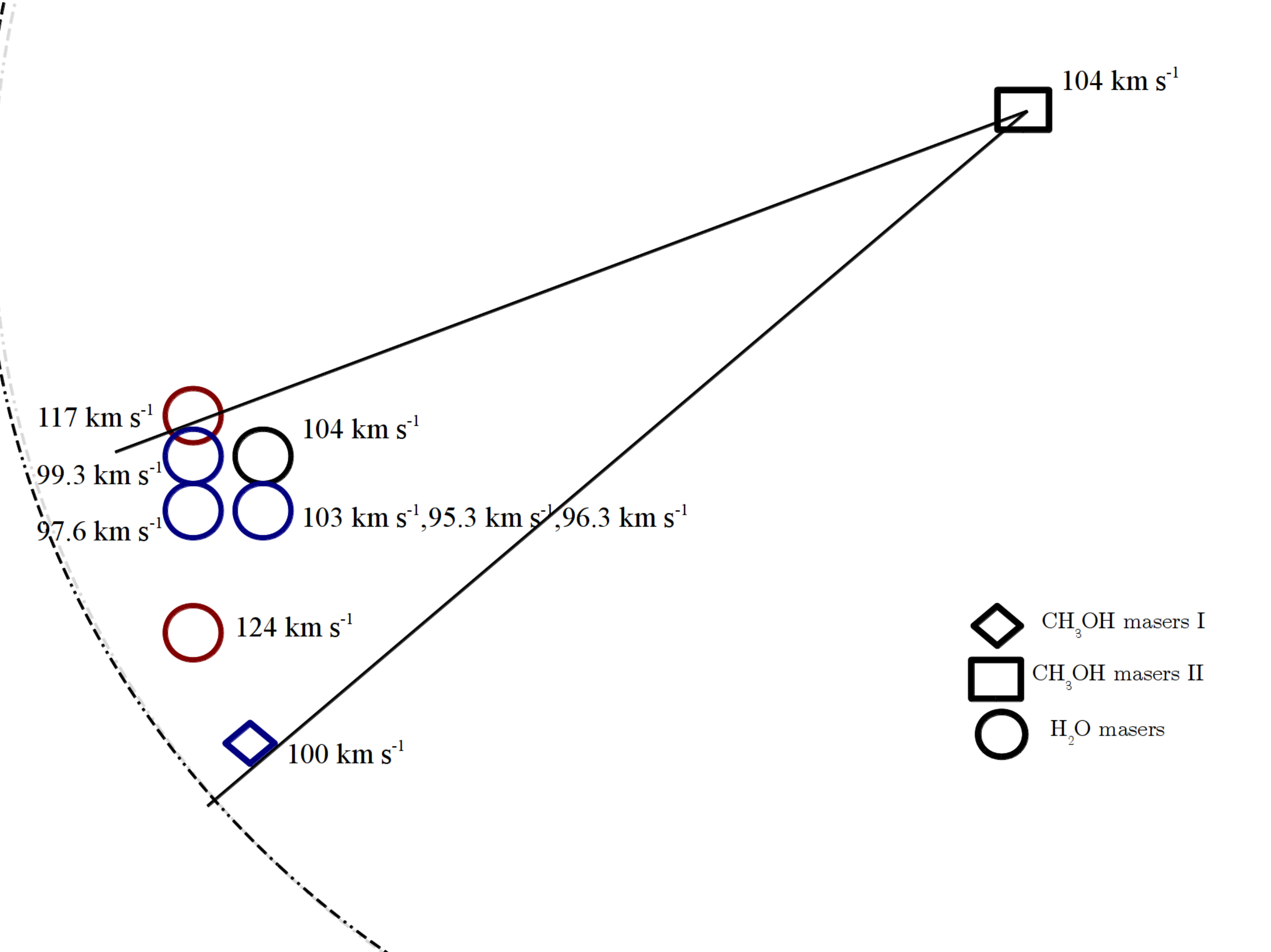}
\caption[Diagram of our disk-outflow model 3 with Class I methanol and H$_{2}$O
masers at the head of the outflow.]{Diagram of our disk-outflow model 3, showing the locations of the
Class II and I methanol masers and the H$_{2}$O masers. We have chosen
to put the information from the G31.41+0.31 star forming region (Figure~\ref{f31.41}
and Table~\ref{t31.41}) in this figure. The symbols are the same
as those used in \S\ \ref{Chapter3}, and are marked at the bottom
right of the figure. Masers that are redshifted with respect to the
Class II methanol maser are shown in red. The dot-dashed line is meant
to be visually suggestive of a bow shock driven into the ambient interstellar
material.}

\label{outflowfig} 
\end{figure}

Our third model is based on the morphology of methanol and H$_{2}$O
masers in G12.20-0.11, G31.41+0.31, and G35.03+0.35 (\S\ \ref{g12.20},
\S\ \ref{g31.41}, \S\ \ref{g35.03}, and \S\ \ref{H2nearClassI}).
Figure~\ref{outflowfig} shows the model of the disk and outflow,
together with the locations of the Class II and Class I methanol masers
and the H$_{2}$O masers. We have chosen to put the information from
G31.41+0.31 (\S\ \ref{g31.41}) in Figure~\ref{outflowfig}. In
G31.41+0.31, 5 H$_{2}$O masers are blueshifted, and 2 H$_{2}$O masers
are redshifted, with respect to the Class II methanol maser (Table~\ref{t31.41}).
The Class I methanol maser is blueshifted with respect to the Class
II methanol maser, and has a very similar $v_{\text{LSR}}$ to the
$v_{\text{LSR}}$ of the H$_{2}$O masers. Therefore, we have put
the Class II methanol maser at the protostar and the Class I methanol
maser along with the blueshifted H$_{2}$O masers in the blueshifted
lobe of an outflow in Figure~\ref{outflowfig}. Based on the position
of the Class I methanol maser and the northern-most blueshifted H$_{2}$O
maser, we get an opening angle for the outflow equal to 22$^{\circ}$,
indicating that the outflow is well collimated. The situation is complicated,
however, by the presence of the two redshifted water masers. Clearly,
they cannot be in the same outflow as the blueshifted water masers.
It is likely, therefore, that they are in the redshifted lobe of a
different outflow. Indeed, there is a UC HII region to the northeast,
and a 4.5 $\mu$m source to the northwest of the redshifted H$_{2}$O
masers (Figure~\ref{f31.41}), and either could be the source of
the redshifted lobe of an outflow.

A similar model works for G35.03+0.35, except that the Class I methanol
and the H$_{2}$O maser are redshifted with respect to the Class II
methanol maser (Table~\ref{t35.03}), so we are seeing the redshifted
lobe of an outflow in this region. Since there is only one H$_{2}$O
maser, we cannot constrain the opening angle of the outflow. The interpretation
is more difficult in G12.20-0.11, where the Class I methanol maser
and two H$_{2}$O masers are redshifted with respect to the Class
II methanol maser, but there are now four blueshifted H$_{2}$O masers
(Table~\ref{t12.20}). Again, it is clear that there must be two
outflows in this region, and if we associate the redshifted H$_{2}$O
masers and Class I methanol maser with the redshifted lobe of one
outflow centered on the protostar near the Class II methanol maser,
the blueshifted H$_{2}$O masers must be in the blueshifted lobe of
another outflow. Indeed, there is a 4.5 $\mu$m source to the east
of the blueshifted water masers, and another one to the southeast
(Figure~\ref{f12.20}), that could be the driving sources of the
blueshifted outflow. The opening angles of the outflows are 11$^{\circ}$
and 19$^{\circ}$.

In summary, all three of these regions suggest a model in which the
Class I methanol maser and the H$_{2}$O masers are located at the
head of outflows where the outflow runs into ambient interstellar
material. However, other models where either the Class I and II methanol
maser, or one type of methanol maser and H$_{2}$O maser, are not
associated with each other, cannot be ruled out.

\subsection{Hybrid Model for G45.47+0.07}

\label{DOM4}

Our 8th source, G45.47+0.07 (\S\ \ref{g45.47} and \S\ \ref{source8})
is likely a hybrid of the disk-outflow model 2 (\S\ \ref{DOM3})
and disk-outflow model 3 (\S\ \ref{DOM2}). The Class I methanol
maser and the H$_{2}$O maser near it are both redshifted with respect
to the Class II methanol maser, so they can be located in the redshifted
lobe of an outflow (disk-outflow model 3; \S\ \ref{DOM2}). Meanwhile,
the Class II methanol maser and the other H$_{2}$O maser could be
in the disk (disk-outflow model 2; \S\ \ref{DOM3}).

\clearpage{}

\section{Conclusion}

\label{Chapter5}

The primary motivation for this paper was to find out whether disk-outflow
systems are compatible with high mass star formation, based on the
existing data on methanol masers, along with other associated tracers
of star formation. We searched the literature for high angular resolution
data on Class I and Class II methanol masers, H$_{2}$O masers, UC
H II regions, and 4.5 $\mu$m infrared sources. We wanted the highest
angular resolution data wherever available, so we looked for VLA (or,
even better, VLBI) data. We started with 44 Class I methanol maser
sources observed with the VLA and listed in \citet{valtts2007}. By
searching the literature for data around these locations, we obtained
high angular resolution data on Class II methanol masers, H$_{2}$O
masers, UC H II regions, and 4.5 $\mu$m infrared sources.

We found a total of thirty regions that contained both Class I and
II methanol masers. For twenty two of these regions high angular resolution
data were only available on the methanol masers. For these regions,
we produced a table of the available data and the projected distances
between the Class I and Class II methanol masers (Table \ref{tlowres}).
For the remaining eight regions, high resolution data were available
on Class I and II methanol masers, water masers, UC H II regions and
4.5 $\mu$m infrared sources. For these eight regions, we prepared
maps of the positions of the Class I and II methanol masers, and the
other star formation tracers listed above (\S\ \ref{g9.62} - \S\ \ref{g75.78}).
We also prepared plots reflecting the center velocities ($v_{\text{LSR}}$)
of the Class I and Class II methanol and H$_{2}$O masers.

Based on these plots, we proposed three disk-outflow models. In all
three models, we placed the Class II methanol maser at the location
of the protostar, and the Class I methanol maser in the outflow. In
our first disk-outflow model (\S\ \ref{DOM1}), the H$_{2}$O masers
are in a linear pattern delineating the outflow. Two of the eight
regions are consistent with this model, although alternative scenarios
cannot be ruled out. In our second disk-outflow model (\S\ \ref{DOM3}),
the H$_{2}$O masers are located in a circumstellar disk near the
Class II methanol maser location; two regions are consistent with
this model. In our third disk-outflow model (\S\ \ref{DOM2}), the
H$_{2}$O masers are located in one or more outflows at the edge of
the shocked region where the outflow interacts with the ambient interstellar
medium; three regions are consistent with this model. Finally, the
8th region is a hybrid of two of these models with one H$_{2}$O maser
located near the Class I methanol maser in the outflow and another
in the circumstellar disk near the Class II methanol maser (\S\ \ref{DOM4}).
All eight regions are therefore compatible with disk-outflow models,
with the Class I methanol maser in the outflow, and the Class II methanol
maser near the protostar.

Together, these models show the utility of coordinated high angular
resolution observations of methanol masers. Since new receivers at
36 GHz are now available at the VLA, there is excellent scope for
unique insight into high mass star formation through future observations
of Class I and Class II methanol masers.
\clearpage{}

\end{document}